\documentclass[10pt]{article}
\usepackage[english]{babel}
\usepackage[a4paper,top=2cm,bottom=2cm,left=3cm,right=3cm,marginparwidth=1.75cm]{geometry}
\usepackage[utf8]{inputenc}
\usepackage{multirow} 
\usepackage{graphicx,float,lipsum}
\usepackage{xcolor}
\usepackage{amsmath, amssymb}
\usepackage{bbold}
\usepackage[colorlinks=true, allcolors=blue]{hyperref}
\usepackage[font=small, labelfont=bf, margin=5mm]{caption}

\title{\bf Scale-invariant random geometry from \\ mating of trees: a numerical study}
\author{\textsc{Timothy Budd}\footnote{Email: \href{mailto:t.budd@science.ru.nl}{t.budd@science.ru.nl}} \qquad \textsc{Alicia Castro}\footnote{Email: \href{mailto:alicia.castro@science.ru.nl}{alicia.castro@science.ru.nl}}\\[3mm]
{\small IMAPP, Radboud University, Nijmegen, The Netherlands.}}
\date{\today}

\begin{document}

\maketitle

\begin{abstract}
\noindent 
The search for scale-invariant random geometries is central to the Asymptotic Safety hypothesis for the Euclidean path integral in quantum gravity.
In an attempt to uncover new universality classes of scale-invariant random geometries that go beyond surface topology, we explore a generalization of the mating of trees approach introduced by Duplantier, Miller and Sheffield.
The latter provides an encoding of Liouville Quantum Gravity on the 2-sphere decorated by a certain random space-filling curve in terms of a two-dimensional correlated Brownian motion, that can be viewed as describing a pair of random trees.
The random geometry of Liouville Quantum Gravity can be conveniently studied and simulated numerically by discretizing the mating of trees using the Mated-CRT maps of Gwynne, Miller and Sheffield.
Considering higher-dimensional correlated Brownian motions, one is naturally led to a sequence of non-planar random graphs generalizing the Mated-CRT maps that may belong to new universality classes of scale-invariant random geometries.
We develop a numerical method to efficiently simulate these random graphs and explore their possible scaling limits through distance measurements, allowing us in particular to estimate Hausdorff dimensions in the two- and three-dimensional setting.
In the two-dimensional case these estimates accurately reproduce previous known analytic and numerical results, while in the three-dimensional case they provide a first window on a potential three-parameter family of new scale-invariant random geometries.
\end{abstract}

\section{Introduction}

According to general relativity, the gravitational force we experience is accounted for by the dynamical geometry of spacetime, as described by a (pseudo-Riemannian) metric on a four-dimensional spacetime manifold satisfying Einstein's classical field equations. 
Since our world appears fundamentally quantum mechanical, general relativity is widely believed to capture merely the low-energy limit of a more fundamental quantum theory of gravity.
A question central to the development of such a theory is: what is to replace the classical smooth metric notion of spacetime geometry?
Proposals for the resulting structure, going under the umbrella term of Quantum Geometry, differ considerably from one approach to the other.
But the characteristics are generally quite different from the smooth metric structure of general relativity because of increasingly large quantum fluctuations at microscopic length scales at and beyond the Planck scale.
This becomes evident when one treats the metric field perturbatively in the gravitational quantum field theory, where its nonrenormalizability spoils the predictive power at microscopic scales.
A scenario in which predictive power can be restored, while retaining the pseudo-Riemannian metric structure as an effective description of spacetime geometry at arbitrarily short length scales, has been proposed in the form of Asymptotic Safety \cite{Weinberg:1980gg,Reuter_Nonperturbative_1998,Kurov:2020csd}.
In this scenario the non-perturbative renormalization group flow of the gravitational quantum field theory approaches an ultraviolet (UV) fixed point at which the dimensionless couplings take finite values and do not change with the energy scale. If these couplings are the ones of geometrical operators, we are led to the conclusion that the quantum laws of the spacetime geometry itself at such a fixed point must be scale-invariant.

Focusing on Euclidean quantum field theories of four-dimensional Riemannian metrics, functional renormalization group methods \cite{Reuter_Nonperturbative_1998} relying on truncations of the renormalization group flow onto finite numbers of couplings have consistently found evidence for the existence of a suitable fixed point, see e.g.\ \cite{Reuter:2019byg}.
From a mathematical or statistical physics point of view, a Euclidean quantum field theory is ideally understood as a probability measure on an appropriate space of field configurations.
This suggest that a mathematical construction of a UV fixed point of Euclidean quantum gravity amounts to the identification of a suitable model of scale-invariant random geometry on the spacetime manifold. 

In the toy model of Euclidean two-dimensional quantum gravity this has been realized in the form of the Brownian plane (and its cousins on compact surfaces, like the Brownian sphere), an exactly scale-invariant probability measure on metric spaces with the topology of the plane, that represents a rigorous construction of what is called the pure-gravity universality class in the physics literature.
It appears not just as the scaling limit of uniform random triangulations of the 2-sphere \cite{le2012scaling}, but also naturally arises from Liouville Quantum Gravity in the absence of matter coupling \cite{ding2021introduction,Gwynne2021,miller2020qle}.
The latter is related to the Liouville Conformal Field Theory \cite{David_Liouville_2016}, which due to the scaling symmetry of conformal field theories makes it clear that we are dealing with a fixed point of the renormalization group flow.
As a consequence of this symmetry, the Brownian plane is not a (random) Riemannian manifold, which would require the geometry in any sufficiently small neighbourhood to resemble that of the Euclidean plane, but is genuinely fractal.
For instance, its Hausdorff dimension, equal to $4$ \cite{legall2007topological}, differs substantially from its topological dimension.

Beyond two dimensions, however, we currently know of no explicit examples of non-trivial scale-invariant random metric spaces with the topology of a three- or four-dimensional manifold.
The construction of natural examples is therefore not only an important ingredient for the asymptotic safety scenario but also presents an important open mathematical problem.
Essentially there are three ways of approaching the problem, mimicking what we know from two dimensions (where all three ways lead to the same results).

The first way would be to construct a scale-invariant quantum field theory on the spacetime manifold that describes the gauge-fixed degrees of freedom of the Riemannian metric. Subsequently, we would need to figure out how to extract the metric geometry from these.
This would be analogous in the two-dimensional setting to first identifying the Liouville Conformal Field theory, which aims to describe two-dimensional Riemannian metrics in conformal gauge, and extract the geometry from there using an appropriate regularization procedure (which has largely been achieved via Quantum Loewner Evolution in the case of pure gravity and Liouville First Passage Percolation in the presence of matter fields).
However, this procedure is difficult to generalize to higher dimensions, because of the lack of good global coordinate gauges and the challenges involved in constructing non-perturbative interacting quantum field theories.

The second way is to introduce discreteness in the field configurations following the philosophy of lattice field theory.
Having a non-zero lattice spacing regularizes the ultraviolet divergences in the path integral while allowing to include field configurations beyond the perturbative regime, as is successfully employed in the numerical investigation of QCD in its strongly-coupled regime. 
The main difference with discretization of matter field theories, in which the lattice provides the geometry on which the fields live, is that in gravity the field should describe the geometry itself. 
This is naturally achieved by allowing the lattice itself to become dynamical, with the gravitational degrees of freedom entirely contained in the combinatorial data describing the lattice and its geometry.
In Dynamical Triangulations \cite{Ambjoern_Diseases_1985,David_model_1985,Ambjoern_Three_1992} the lattice is constructed by gluing equilateral simplices (triangles, tetrahedra, \ldots, depending on the dimension in which the model is considered), while much more general random planar map models have been investigated in two dimensions.
In order to find a scale-invariant random geometry, representing a potential Euclidean quantum field theory at the UV fixed point, it is necessary to take a scaling limit where the number of building blocks is taken to infinity while their size is taken to zero. 
As is well known from statistical physics, for such a non-trivial scaling limit to exist, the discrete model must be critical, in that it exhibits diverging correlation lengths.
Besides criticality, another very important criterion in the case of random geometry is that the manifold structure does not degenerate in the scaling limit. For instance, in Dynamical Triangulations of the 3-sphere, the piece-wise flat geometries built from equilateral tetrahedra have the topology of the 3-sphere and display criticality in the so-called branched polymer phase of the model.
However, numerical simulations indicate that shrinking the building blocks leads to the topology degenerating into that of trees, nothing like the manifold structure of the 3-sphere.
Apart from the branched polymers, simulations of Dynamical Triangulations in three and four dimensions have not (yet) uncovered critical phenomena that escape this branched-polymer universality.
This means that the lattice approach is yet to uncover concrete opportunities to establish scale-invariant random geometries on three- and four-dimensional manifolds.
There are however promising avenues in models that restrict the family of triangulations considered.
Simulations suggest that four-dimensional \emph{Causal} Dynamical Triangulations (CDT) \cite{loll2019quantum} feature continuous phase transitions \cite{ambjorn2017characteristics,ambjorn2020higher} where one expects criticality to be found, while a numerical investigation of a recently proposed model of three-dimensional dynamical triangulations assembled from triples of trees is underway \cite{budd2022threespheres}.

The third route towards scale-invariant random geometry, the one that we follow in this work, also aims to assemble geometries out of simpler building blocks, but instead of relying on criticality and scaling limits to approach scale-invariance with random discrete objects, one takes the building blocks themselves to be scale-invariant. 
If the assembly does not spoil the scale-invariance and the resulting geometry has the desired topology, this provides a very economical way of uncovering new universality classes.
A simple example of such an assembly procedure is that of the Continuum Random Tree (CRT), which is the scaling limit of the branched polymer universality class mentioned above, out of Brownian motion \cite{AldousCRT}.
Brownian motion, seen as a random continuous real function on the line (or equivalently as a massless free scalar field in one dimension), is the prime example of a scale-invariant random object.
This random real function naturally gives rise to a gluing procedure of the real line into a topological tree with a metric, namely the CRT, that shares the same scaling symmetry (see Section~\ref{subsection:matedcrt} for a discussion).
Since the CRT is not a topological manifold, its relevance for quantum gravity is not obvious, but stems from the possibility of using the CRT itself as building block for larger random geometries.

The hope of assembling manifolds out of random trees may seem far-fetched, but is well established in two-dimensional quantum gravity. 
At the level of discretized surfaces, bijective encoding of planar maps in treelike combinatorial structures has a long tradition, starting with Mullin's bijection for tree-decorated maps \cite{mullin_1967} and the Cori--Vauquelin--Scheafer bijection between quadrangulations \cite{cori1981planar,Schaeffer1998ConjugaisonDE} and labeled trees.
The study of the latter bijection in the scaling limit, in which the discrete trees approach the CRT, paved the way for a mathematically rigorous construction of (and convergence to) the scale-invariant Brownian sphere \cite{marckert2006limit,le2012scaling,miermont2013brownian}.
Generalizations of Mullin's bijection to random planar maps decorated by various critical statistical systems \cite{sheffield2016quantum,bernardi2007bijective,Kenyon_Bipolar_2019,Li_Schnyder_2017} hinted at a different appearance of CRTs in the continuum limit, in a way that ties in closely with Liouville Quantum Gravity and conformally invariant random curves described by Schramm--Loewner Evolution (SLE).
In the foundational paper \cite{duplantier2014liouville} by Duplantier, Miller and Sheffield this approach, going under the name of Mating of Trees, was introduced in its generality.
Starting from two independently sampled CRTs, there exists a simple assembly procedure that identifies points in the contours of the two trees, resulting in a scale-invariant random metric space that (almost surely) has the topology of the 2-sphere.
This metric space is known to correspond with Liouville Quantum Gravity for a particular value of its coupling constant ($\gamma=\sqrt{2}$). 
Remarkably, any other value of this coupling $\gamma\in(0,2)$ associated to a gravitational universality class in the presence of an arbitrary matter conformal field theory (as long as the matter central charge is below $1$), can be achieved by introducing a correlation between the pair of CRTs.
As alluded to above, a CRT is assembled from a Brownian motion, meaning that a pair of CRTs is naturally obtained from the two coordinates of a two-dimensional Brownian motion, and this correlation can be understood as the choice of a non-trivial covariance matrix for the latter Brownian motion. 

The fact that the CRT provides the universal building block for essentially all scale-invariant random geometries relevant to two-dimensional quantum gravity naturally raises the question whether higher-dimensional random geometries can be constructed in similar fashion.
One can question whether this is sufficiently motivated from a path integral perspective on quantum gravity, but given that at present we do not know of a single explicit example of a scale-invariant random geometry with three-dimensional manifold topology one should not set too stringent conditions.
In this work we propose a rather straightforward generalization of the mating of trees construction, in which the pair of correlated CRTs is replaced by a triple. If the result has a well-defined and scale-invariant random metric structure, something that requires checking, it necessarily gives rise to new universality classes beyond random surfaces.
Naturally the model possesses a three-dimensional parameter space as opposed to the one-dimensional parameter space of mated-CRT surfaces.
One of the critical exponents, the string susceptibility, can be calculated (analytically for special regions and numerically elsewhere) and displays a non-trivial dependence on the three parameters, suggesting that they really parameterize an entire family of new universality classes.
To start exploring the parameter space we develop a numerical toolbox to simulate the result of mating a triple of trees and measure an important critical exponent related to the metric: the Hausdorff dimension, which governs the relative scaling between volume and radius of geodesic balls in the geometry. 
Whether the topology induced by the metric really is that of a three-dimensional manifold requires a more refined analysis that is beyond the scope of the current work.

This paper is organised as follows: we start in Section~\ref{section2} by reviewing the mating of trees approach to two-dimensional quantum gravity and its relation to random planar map models and Liouville Quantum Gravity. To access the metric properties of mated CRTs, it is necessary to consider regularizations in the form of Mated-CRT maps, where geodesic distances can be conveniently approximated by graph distances on the map. In Section~\ref{sec:matedcrtgeneralized} we propose a generalization of the Mated-CRT maps to Mated-CRT graphs associated to Brownian motion of arbitrary dimension and outline an algorithm to effectively sample them. 
Section~\ref{section_numerical} describes our numerical implementation and simulation of Mated-CRT graphs associated to two- and three-dimensional Brownian motions and a finite-size scaling analysis of geodesic distances.
The resulting estimates of the Hausdorff dimension in the two-dimensional case serve as an important benchmark of the numerical method, while in three dimensions the estimated Hausdorff dimensions combined with the computed string susceptibilities provide a first window into a large family of potentially three-dimensional random geometry universality classes. In the last section we discuss our results in the context of scale-invariant random geometries and Quantum Gravity.

\subsection*{Acknowledgments}

This work is part of the START-UP 2018 programme with project number 740.018.017, which is financed by the Dutch Research Council (NWO). TB also acknowledges support from the VIDI programme with project number VI.Vidi.193.048, which is financed by the Dutch Research Council (NWO).


\section{Mating of trees: 2D quantum gravity from Brownian motion}\label{section2}

The basic idea of the Mating of Trees approach is that both the geometry and the matter degrees of freedom in two-dimensional quantum gravity on the 2-sphere can be encoded in a single continuous path in the Euclidean plane.
For a detailed story of the correspondence, we direct the mathematically-inclined audience towards the foundational paper by Duplantier, Miller and Sheffield \cite{duplantier2014liouville} and the recent survey by Gwynne, Holden and Sun \cite{Gwynne_Mating_2019}, as well as the many references in the latter.
In this section we provide a high-level introduction to the topic for those that are not entirely comfortable with the probability theory literature.  
First, we will review two examples of mating-of-trees bijections between discrete surfaces decorated with statistical systems and certain discrete walks in the quadrant. 
Next, we will explain how this picture extends to the continuum limit and generalizes to the full family of two-dimensional quantum gravity theories coupled to conformal matter. 

\subsection{First discrete example: spanning-tree decorated quadrangulations}\label{sec:stquad}

\begin{figure}[H]
  \centering
    \includegraphics[width=0.85\textwidth]{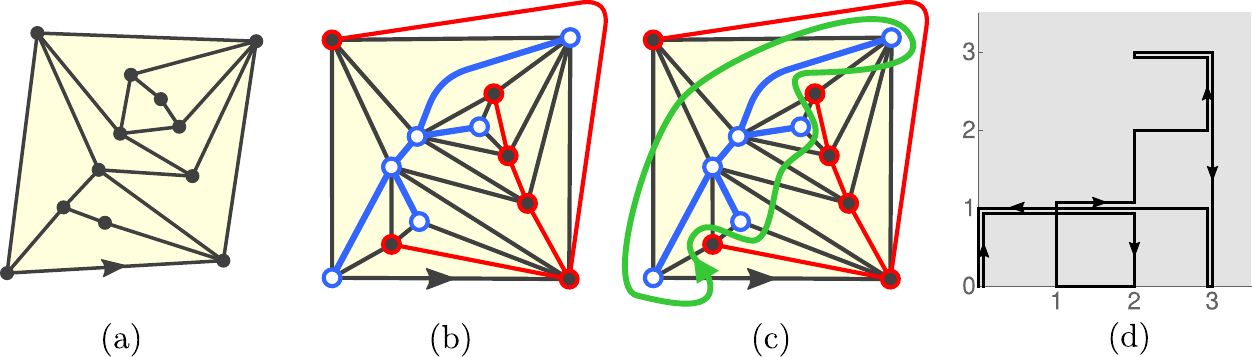}
  \caption{(a) A rooted quadrangulation (note that the white outer region is a face of degree four as well). (b) A spanning-tree-decorated quadrangulation. (c) The space-filling curve. (d) The corresponding excursion $(Z_i)$ in the quadrant. Figure adapted from \cite{Barkley_2019}.
  \label{fig:spanningtreealgorithm}}
\end{figure}

The simplest example of a model of discrete surfaces that is naturally encoded by a walk in the quadrant, is that of spanning-tree decorated quadrangulations, which goes back to a bijection of Mullin \cite{mullin_1967} in the sixties. 
To define the model, we need to introduce some terminology.
A \emph{planar map} is a planar graph, in which loops and multiple edges between vertices are allowed, together with a proper embedding in the 2-sphere.
A region in the sphere that is delimited by edges of the map is called a \emph{face} and the \emph{degree} of a face is the number of edges in its contour.
A planar map is said to be \emph{rooted} if it has a distinguished oriented edge.
A \emph{quadrangulation} is a planar map in which all faces have degree four (Figure \ref{fig:spanningtreealgorithm}a).
To see that a quadrangulation describes a discrete surface, it is sometimes useful to think of the faces as identical unit squares equipped with the Euclidean metric and the incidence relations of the map as prescriptions on how to glue these squares along their sides in order to obtain a piecewise flat metric on the sphere.

A \emph{spanning-tree-decorated quadrangulation} is a quadrangulation together with a choice of diagonal in each face, such that the graph formed by the diagonals alone has no loops (Figure \ref{fig:spanningtreealgorithm}b).
In this case the diagonals necessarily take the form of two disjoint trees that together span the vertices of the quadrangulation, hence the name of the model.
The two vertices of the root edge belong to the two different trees and mark a root for each of them.
If one assigns equal Boltzmann weight to each spanning-tree-decorated quadrangulation (in other words, one samples uniformly), one may think of this decoration as a statistical system coupled to the geometry of the surface described by the quadrangulation, thus as a rather abstract form of matter.
Note that the presence of the statistical system has an (entropic) effect on the geometry, as the number of decorations differs from one quadrangulation to the other.

According to Mullin \cite{mullin_1967} rooted spanning-tree-decorated quadrangulations with $n$ faces are in bijection with excursions in the quadrant of length $2n$ with unit steps parallel to the axes.
An \emph{excursion in the quadrant} is a walk $Z_0,Z_1,\ldots,Z_{2n} \in \mathbb{Z}_{\geq0}^2$ with $2n$ steps that starts and ends at the origin, $Z_0 = Z_{2n}=0$ (Figure \ref{fig:spanningtreealgorithm}d).
The bijection is rather easy to understand: there exists a unique closed path on the surface starting and ending at the root edge that intersects all edges of the quadrangulation while avoiding all diagonals (Figure \ref{fig:spanningtreealgorithm}c).
The corresponding excursion simply records for the $i$th visited edge the heights $Z_i \in \mathbb{Z}_{\geq0}^2$ of its left and right extremity in the tree, where the \emph{height} of a vertex in a tree is the distance in the tree to its root.
From the figure it should be clear that between consecutive visits, exactly one of the heights changes by $\pm1$, so $Z_{i+1}-Z_i \in \{(0,\pm1),(\pm1,0)\}$ and one indeed obtains the desired excursion in the quadrant. 
It is straightforward to check that any such excursion can be obtained in this was and that the quadrangulation together with its decoration can be reconstructed from the excursion. 

In light of what follows, it is useful to think of the reconstruction starting from an excursion as a three step procedure.
In the first step one only examines the sequence of horizontal steps of the excursion (and ignoring the vertical coordinate), which encodes a \emph{Dyck path}, i.e. a walk with unit steps on the non-negative integers starting and ending at zero. 
It encodes a plane tree, which we draw in blue.
In the second step, a red tree is constructed similarly from the vertical steps of the excursion.
Now every visit of the excursion naturally corresponds to a pair of corners, one on each tree, such that following the excursion corresponds to tracing the contour of the blue tree in counterclockwise direction and the contour of the red tree in clockwise direction.
Finally, in the third step the blue and red tree are ``mated'' into a spanning-tree-decorated quadrangulation by drawing a black edge between each pair of corners. 

\subsection{Second discrete example: site-percolated triangulations}\label{sec:perctri}

\begin{figure}[H]
  \centering
    \includegraphics[width=\textwidth]{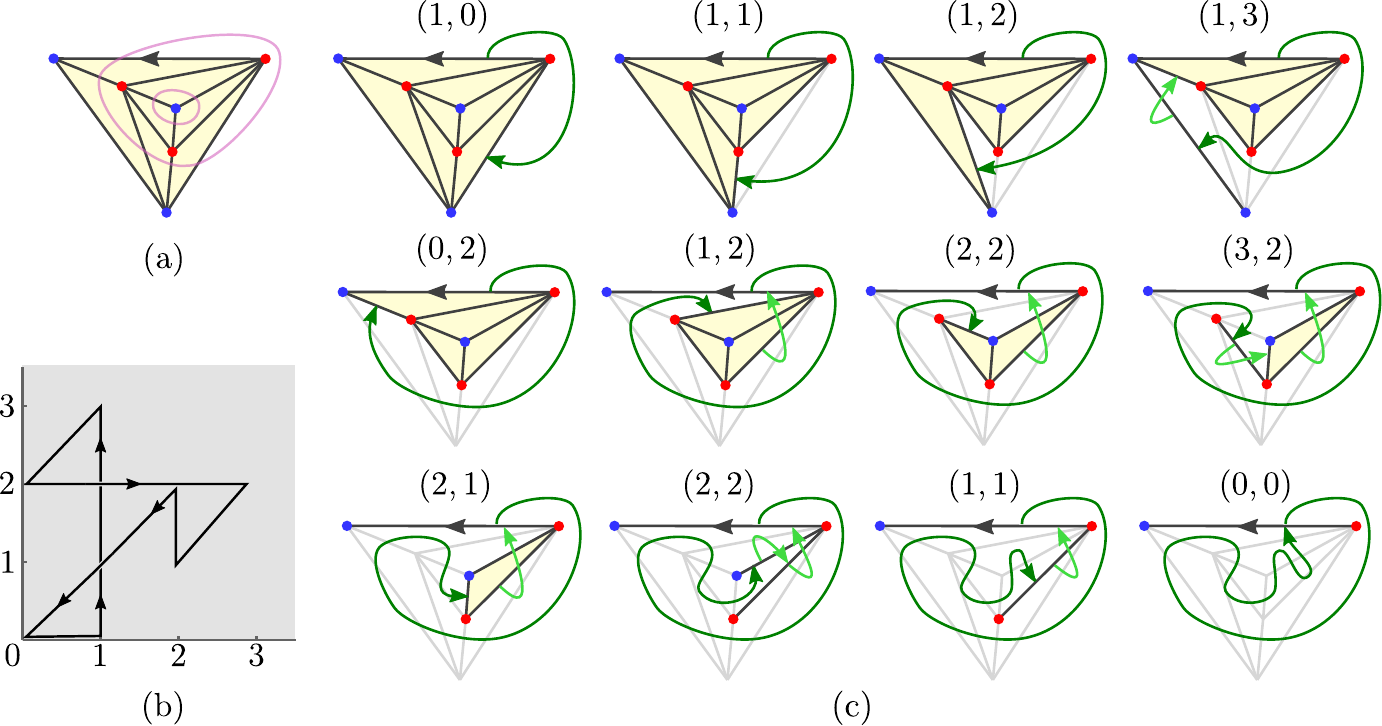}
  \caption{(a) A rooted site-percolated triangulation with cluster interfaces indicated in pink. (b) The corresponding excursion in the quadrant with increments $(1,0)$, $(0,1)$ and $(-1,-1)$. (c) The peeling exploration. The dark green curve illustrates the iteratively constructed exploration path, while the lighter green curves indicate the required detours. The integers above each map are contour lengths on the left respectively right of the exploration path.
  \label{fig:percolationalgorithm}}
\end{figure}

The previous example is archetypal, where it is intuitively clear that the surface can be encoded in trees, because the decoration already takes the form of a tree.
Admittedly, it is not the most natural statistical system one would think of when trying to couple quantum gravity to a matter field.
So let us look at a model that has a simpler interpretation, but for which the trees are well hidden.
This is the model of (loopless) triangulations with site percolation \cite{angel2005scaling,bernardi2007bijective,gwynne2021joint}.
A \emph{triangulation} is a planar map in which all faces have degree three and it is \emph{loopless} if it has no edges starting and ending on the same vertex.
A \emph{site percolation} on a rooted triangulation is simply an assignment of one of two colors, say blue and red, to each vertex of the map, with the only requirement that the root edge points from a red to a blue vertex (see Figure \ref{fig:percolationalgorithm}a).
If one assigns equal Boltzmann weight to every such rooted site-percolated triangulation with $2n$ triangles, we obtain a very simple example of a statistical system on a random surface.
One could think of this system as the high-temperature limit of the standard Ising model living on the vertices of the triangulation.
Note that, contrary to the spanning-tree-decorated quadrangulations, each triangulation admits the same number of distinct site percolations, namely $2^{n}$ because there are precisely $n$ vertices that are not incident to the root edge.  
This means that the statistical system does not affect the statistics of the geometry, and we are dealing with a model whose geometry lives in the universality class of pure gravity, coupled to a rather trivial type of matter in the form of white noise.

Just like in the previous example, we would like to find a self-avoiding closed curve that in some sense explores the full triangulation.
It is natural to consider the partition of the vertex set into monochromatic clusters and examine the cluster interfaces, which naturally correspond to a collection of disjoint closed loops on the graph dual to the triangulation (Figure \ref{fig:percolationalgorithm}a).
Since the root edge crosses an interface, it is natural to start the exploration along this interface.
Unless the site-percolated triangulation is exactly of the type of Figure \ref{fig:spanningtreealgorithm}b, with exactly one blue cluster and one red cluster and such that the monochromatic edges in each cluster form a tree, the exploration will return to the root edge before having explored the full map.
The rough idea described in \cite{bernardi2007bijective} is that one can merge all cluster interfaces into a single exploration by following an interface and taking detours into neighbouring interfaces at the very last opportunity before they become inaccessible. 

\begin{figure}[H]
  \centering
    \includegraphics[width=\textwidth]{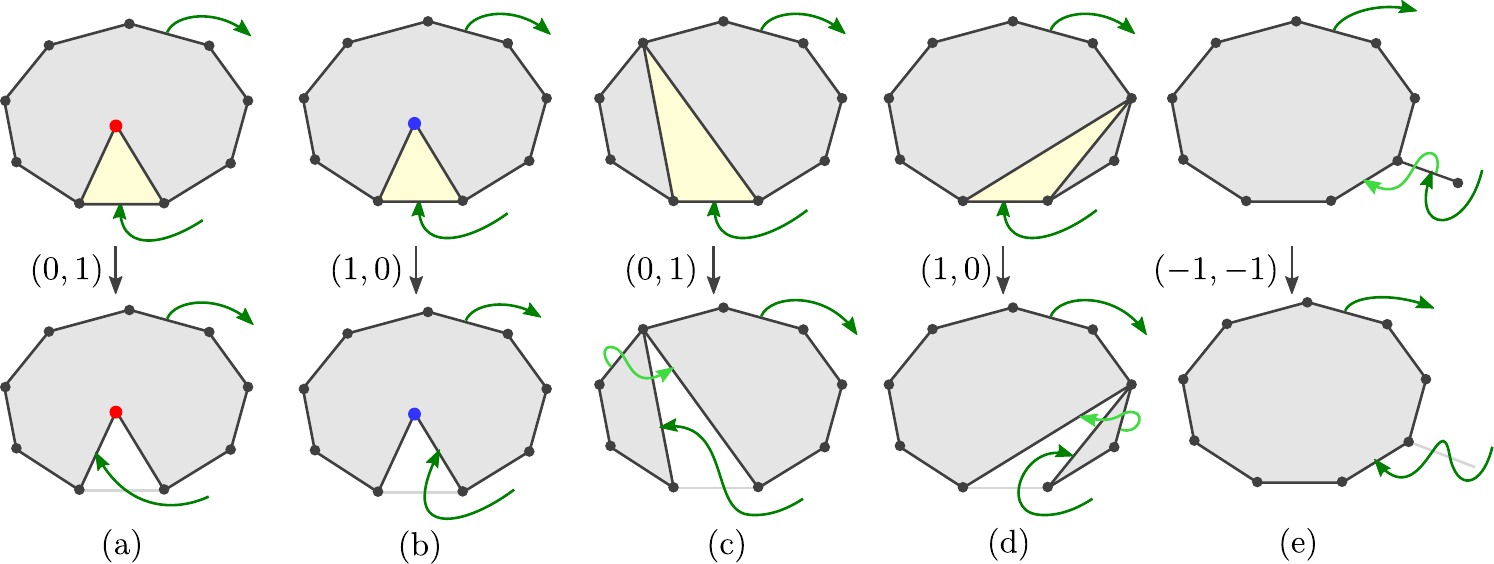}
  \caption{The five types of possibilities when peeling away an edge in the exploration. The pair of integers indicates the length change of the contour to the left respectively right of the exploration path.
  \label{fig:percolationmoves}}
\end{figure}

More precisely, one may setup a peeling exploration that visits all $3n$ edges of the triangulation as follows (see \cite[Section 2.3]{gwynne2021joint}).
We start the exploration in the triangle at the right of the root edge and position the tip of the exploration at the other non-monochromatic edge of that triangle (see the top left of Figure \ref{fig:percolationalgorithm}c).
Then at each step the edge $e$ at the tip is removed, in such a way that we eventually return to the root edge (the final \emph{target}).
There are five different cases (a to e) to be considered, which are summarized in Figure \ref{fig:percolationmoves}. 
If $e$ is adjacent to a triangle containing a non-boundary vertex, then the exploration traces the cluster interface, meaning that it turns left or right depending on the color of the vertex (cases a and b). 
If however, all vertices of the triangle are on the boundary, one considers the two components that are separated by the triangle, and one implements a detour through the component that does not contain the target.
The choice of detour is shown in Figures \ref{fig:percolationmoves}c and d by a light green arrow pointing from one component, where it can be regarded as an intermediate target for the exploration, to the other, indicating where the exploration will continue after the first component has been fully explored.
Finally, if instead of a triangle the edge $e$ is adjacent to a detour, the detour is followed (case e).
An example of a full exploration is shown in Figure \ref{fig:percolationalgorithm}c.

To obtain a lattice walk, at every step $i=1,\ldots,3n$ of the exploration one keeps track of the distances $Z_i \in \mathbb{Z}_{\geq 0}^2$ in clockwise respectively counterclockwise direction along the contour between the tip of the exploration and the root edge.
It is easily seen that the only possible changes in these distances are $(0,1)$, $(1,0)$ and $(-1,-1)$, so one obtains an excursion with these increments in the quadrant of length equal to the number of edges of the triangulation (Figure \ref{fig:percolationalgorithm}b). 
These walks are known as Kreweras walks \cite{Kreweras1965}.
It is a non-trivial fact that this determines a bijection, see \cite[Theorem 2.2]{gwynne2021joint} and the earlier references \cite{bernardi2007bijective,Bernardi_Percolation_2018}.

\subsection{Scaling limit of the walks}\label{sec:walkscaling}

A first consequence of these bijections is that one can easily understand the asymptotics of the enumeration.
Let us denote by $\mathcal{Z}^*_n$ the canonical partition function with unit Boltzmann weight per configuration, meaning simply the total number of decorated rooted planar maps in the model.
In the case of spanning-tree decorated quadrangulations, the number of simple excursions of length $2n$ is easily found to be 
\begin{equation}\label{eq:Zspanning}
  \mathcal{Z}_n^{\text{spanning-tree}} = \operatorname{Cat}(n)\operatorname{Cat}(n+1) \stackrel{n\to\infty}{\sim} \frac{4}{\pi}\,4^{2n} n^{-3},
\end{equation}
where $\operatorname{Cat}(n) = \frac{1}{n+1}\binom{2n}{n}$ are the Catalan numbers.
Note that the exponential growth $4^{2n}$ reflects the four different increments available for each of the $2n$ steps of a simple walk.
In the case of site-percolated triangulations, the number of Kreweras excursions of length $3n$ is \cite{Kreweras1965,bernardi2007bijective}
\begin{equation}\label{eq:Zpercolation}
  \mathcal{Z}_n^{\text{percolation}} = \frac{4^n}{(n+1)(2n+1)} \binom{3n}{n} \stackrel{n\to\infty}{\sim} \sqrt{\frac{3}{16\pi}}\,3^{3n} n^{-5/2}.
\end{equation}
Also here the $3^{3n}$ agrees with the three possible increments of the walk at each of the $3n$ steps.
More importantly, the exponents of the power-law correction differ between the two models. 

This should not be surprising, because the models feature qualitatively different matter systems and should be expected to belong to different universality classes.
In general, for a model of random geometry the partition function is expected to scale with $n$ as
\begin{equation}
  \mathcal{Z}^*_n\overset{n\rightarrow\infty}{\sim}Cn^{\gamma_\mathrm{s}-2}\kappa^n,\label{numberofgraphs}
\end{equation}
where $\gamma_\mathrm{s}$ is a critical exponent, known in the physics literature as the \emph{string susceptibility}.
If the universality class corresponds to two-dimensional quantum gravity coupled to a matter conformal field theory with central charge $c \in (-\infty,1]$, then the KPZ formula predicts 
\cite{Knizhnik:1988ak}
\begin{equation}
   \gamma_\mathrm{s}=\frac{c-1-\sqrt{(c-1)(c-25)}}{12}.
    \label{gamma_string_with_c}
\end{equation}
We see that spanning-tree-decorated quadrangulations feature the exponent $\gamma_\mathrm{s}=-1$ corresponding to $c=-2$ and the site-percolated triangulations exponent $\gamma_\mathrm{s}=-1/2$ corresponding to $c=0$, in accordance with the expectation that the latter live in the pure gravity universality class.

Why do we see two different exponents appearing in the enumerations of excursions in the quadrant?
In the large-$n$ limit the random walks in the quadrant, rescaled by $1/\sqrt{n}$, approach a Brownian excursion in the quadrant, i.e. a two-dimensional Brownian motion that is conditioned to start and end at the origin and remain in the quadrant.
In order to specify this process, it is sufficient to know the covariance of the unrestricted Brownian motion, which appears as the limit of the unrestricted random walks.
It is precisely in this covariance that the two models differ.  
Indeed, denoting the $x$ and $y$ components of a walk on $\mathbb{Z}^2$ by $L_t$ and $R_t$ respectively, an unrestricted simple random walk on the square lattice has covariance
\begin{equation}
  \operatorname{Var}(L_t) = \operatorname{Var}(R_t) = \frac{t}{2},\qquad \operatorname{Cov}(L_t,R_t) = 0,
\end{equation}
while an unrestricted Kreweras random walk satisfies 
\begin{equation}
  \operatorname{Var}(L_t) = \operatorname{Var}(R_t) = \frac{2t}{3},\qquad \operatorname{Cov}(L_t,R_t) = \frac{t}{3}.
\end{equation}
In general, for a random walk with
\begin{equation}\label{eq:covrho}
  \operatorname{Cov}(L_t,R_t) = \rho \operatorname{Var}(L_t) = \rho \operatorname{Var}(R_t), \qquad \rho \in (-1,1)
\end{equation}
it is known \cite{denisov2015random} that the number of excursions in the quadrant of length $n$ grows like 
\begin{equation}
  C\, n^{-1-\frac{\pi}{\arccos(-\rho)}}\,\kappa^n,
\end{equation}
for some $C>0$ and $\kappa>1$.
Note that for $\rho=0$ respectively $\rho=1/2$ this indeed agrees with \eqref{eq:Zspanning} respectively \eqref{eq:Zpercolation}.
Moreover, it suggests that any other model of random decorated planar maps that admits a bijection with walks in the quadrant and belongs to a universality class with a certain central charge $c$ must satisfy $\rho = -\cos\left(\frac{\pi}{1-\gamma_\mathrm{s}}\right)$.
Several further examples are indeed known for which this is the case, including bipolar-oriented triangulations ($c=-7$, $\gamma_\mathrm{s}=-2$, $\rho = -1/2$) \cite{Kenyon_Bipolar_2019} and Schnyder-wood-decorated triangulations ($c=-25/2$, $\gamma_\mathrm{s}=-3$, $\rho = -1/\sqrt{2}$) \cite{Li_Schnyder_2017}.

\subsection{Mating of trees and Liouville Quantum Gravity}

The discrete examples make one wonder whether there is a continuum interpretation to the encoding by trees and whether it extends to other universality classes of two-dimensional quantum gravity coupled to conformal matter with $c\in(-\infty,1]$.
This has indeed been shown to be the case in a framework going under the name of mating of trees \cite{duplantier2014liouville}, putting the case $c=1$ aside with its peculiarities \cite{Aru2021}.
In order to understand the result, we need to explain first how to describe geometry and space-filling curves in the continuum.
We are dealing with quantum gravity on the 2-sphere, which is conveniently represented by the Riemann sphere $\hat{\mathbb{C}} = \mathbb{C} \cup \{\infty\}$.
Let $\hat{g}_{ab}$ be some fixed conformal\footnote{Conformal in the sense that $\hat{g}_{ab}$ preserves the angles of $\hat{\mathbb{C}}$, and thus corresponds to a metric of the form $f(z)(\mathrm{d}x^2+\mathrm{d}y^2)$ for $z=x+iy\in\hat{\mathbb{C}}$ for some function $f:\hat{\mathbb{C}}\to[0,\infty)$. A natural choice, corresponding to constant curvature $\hat{R}=8\pi$ and unit area, is $f(z) = \frac{1}{\pi}(1+|z|)^{-2}$.} background metric on $\hat{\mathbb{C}}$ of unit area.
Then, we are after a random real field $\phi$ on $\hat{\mathbb{C}}$, that we informally interpret as describing a random Riemannian metric $g_{ab} = e^{\gamma \phi} \hat{g}_{ab}$ of unit volume, and independently a random continuous space-filling curve $\eta_{\hat{g}} : [0,1] \to \hat{\mathbb{C}}$ such that $\eta_{\hat{g}}(0) = \eta_{\hat{g}}(1)$ and $\eta_{\hat{g}}([s,t])$ has volume $t-s$ with respect to the background measure $\sqrt{\hat{g}}\,\mathrm{d}^2z$ for $0<s<t<1$. 
The former is provided by Liouville Quantum Gravity (LQG) and the latter by Schramm--Loewner Evolution (SLE), which we both briefly discuss.

By the uniformization theorem, any two-dimensional Riemannian metric on the sphere is isometric to a conformal rescaling $g_{ab} = e^{\gamma \phi}\hat{g}_{ab}$ of the background metric $\hat{g}_{ab}$. 
Liouville Quantum Gravity with coupling constant $\gamma \in (0,2)$ is the path integral quantization of this field $\phi$ (known as the dilaton) with action 
\begin{equation}
    S_{L}=\frac{1}{4\pi}\int d^2x \sqrt{\hat{g}}\left(\hat{g}^{ab}\partial_a\phi\partial_b\phi +Q\hat{R}\phi + 4\pi\hat{\mu} e^{\gamma\phi}\right),
\end{equation}
where
\begin{equation}\label{eq:Q_gamma}
    Q=\frac{\gamma}{2}+\frac{2}{\gamma},
\end{equation}
$\hat{R}$ is the scalar curvature of $\hat{g}_{ab}$ and $\hat{\mu} > 0$ a parameter known as the cosmological constant.
If gravity is coupled to a conformal matter field with central charge $c \in (-\infty, 1)$, then the parameter $Q\in(2,\infty)$ is related to $c$ via
\begin{equation}
    c=25-6 Q^2.\label{conformal_charge}
\end{equation}
It is not obvious that one can make sense of a random field with density proportional to $e^{-S_L}$ in an appropriate space of generalized functions.
Luckily $\phi$ is closely related to the Gaussian Free Field (GFF) on $\hat{g}_{ab}$, the free massless real scalar field with action $S_L$ in which $Q$ and $\hat{\mu}$ are set to zero, which has an unambiguous probabilistic interpretation if we restrict the constant mode, for instance, by requiring zero mean (see \cite{Werner2020} for a recent introduction to the GFF).
Accounting for the residual M\"obius symmetry by marking three points, say $z_1,z_2,z_3\in \hat{\mathbb{C}}$, it can be shown that $\phi$ is obtained from the GFF by a deterministic position-dependent shift \cite{David_Liouville_2016,Aru_Two_2017}.

\begin{figure}
  \centering
  \includegraphics[width=\linewidth]{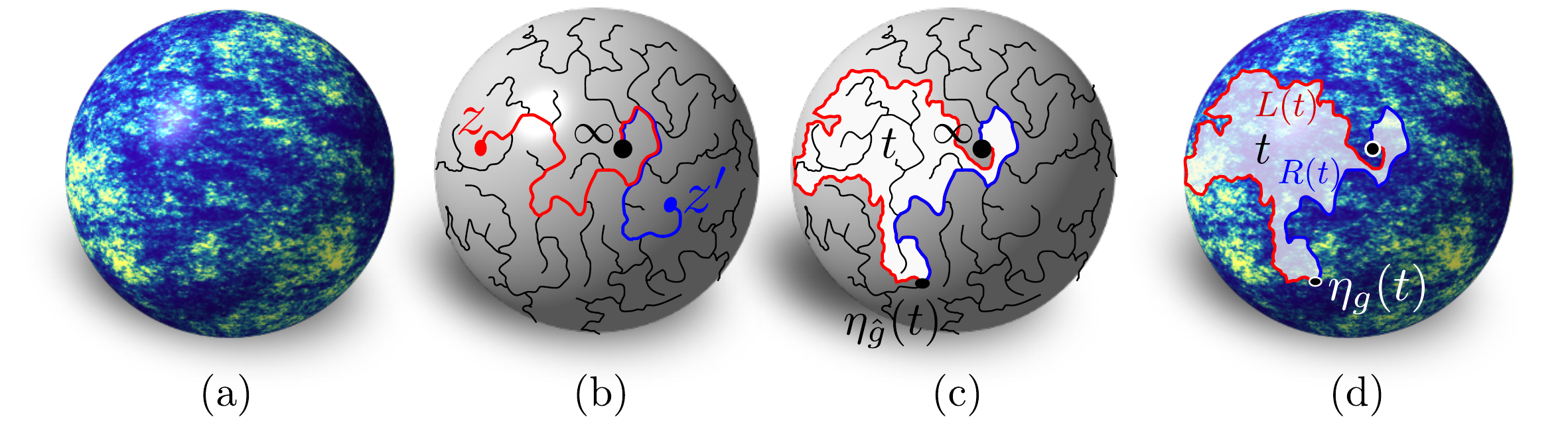}
  \caption{(a) LQG$_\gamma$: simulation of the random measure $\mu_\phi$ on the round 2-sphere (lighter regions contain more quantum area than darker regions). (b) Space-filling SLE$_{\kappa'}$: an illustration of the imaginary geometry flow lines from $z$ and $z'$  to $\infty$. Here $\eta_{z}$ (in red) meets $\eta_{z'}$ (in blue) from the left, and thus $z$ precedes $z'$ in the space-filling curve. (c) $\eta_{\hat{g}}([0,t])$ is a closed region (in white) of $\hat{g}$-area $t$ in $\hat{\mathbb{C}}$ with $\infty$ and $\eta_{\hat{g}}(t)$ on its boundary. (d) LQG$_\gamma$ + SLE$_{\kappa'}$: the region $\eta_{g}([0,t])$ has quantum area $\mu_\phi(\eta_{g}([0,t]))=t$ and $L(t)$ and $R(t)$ are the left and right boundary lengths of this region measured by $\nu_\phi$ respectively. \label{fig:lqg-sle}}
\end{figure}

The resulting random field $\phi$ is not defined point-wise but should be viewed as a generalized function.
In order to make sense of the rescaling $g_{ab}=e^{\gamma \phi}\hat{g}_{ab}$, it is thus necessary to consider a suitable regularization.
For instance, we could look at the circle average $\phi_\epsilon(z)$, taken to be the average of $\phi(z)$ over a circle of radius $\epsilon$ around $z$.
Then, a normalized \emph{quantum area} measure on $\hat{\mathbb{C}}$ can be defined via
\begin{equation}
    \mu_\phi=\lim_{\epsilon \to 0} \frac{e^{\gamma \phi_\epsilon(z)}\sqrt{\hat{g}}\,\mathrm{d}^2z}{\int_{\hat{\mathbb{C}}} e^{\gamma \phi_\epsilon(z)}\sqrt{\hat{g}}\,\mathrm{d}^2z},\label{LQG_area_measure}
\end{equation}
that is independent of $\hat{g}_{ab}$ and is such that $z_1,z_2,z_3$ are uniform points for $\mu_\phi$\footnote{More precisely, this means that if we sample three new independent points $z_1',z_2',z_3'$ with probability distribution $\mu_\phi$ and look at the transformation of $\mu_\phi$ under the unique M\"obius transformation that sends $z_i'$ to $z_i$, $i=1,2,3$, then the result has the same law as $\mu_\phi$.}.
See Figure~\ref{fig:lqg-sle}a for an illustration.
Similarly, one may introduce a \emph{quantum length} measure via
\begin{equation}
    \nu_\phi=\lim_{\epsilon \to 0} \frac{e^{\gamma \phi_\epsilon(z)/2}\sqrt[4]{\hat{g}}\,|\mathrm{d}z|}{\sqrt{\int_{\hat{\mathbb{C}}} e^{\gamma \phi_\epsilon(z)}\sqrt{\hat{g}}\,\mathrm{d}^2z}}.\label{LQG_length_measure}
\end{equation}
Given a region $U \subset \hat{\mathbb{C}}$ and a curve $\Gamma : [0,1] \to \hat{\mathbb{C}}$, $\mu_\phi(U)$ and $\nu_\phi(\Gamma([0,1]))$ should be interpreted as rigorous definitions for the usual area $\int_{U} \sqrt{g}\mathrm{d}^2z$ and length $\int_0^1 \sqrt[4]{g}|\Gamma'(t)|\mathrm{d}t$ as measured by the Riemannian metric $g_{ab}=e^{\gamma \phi}\hat{g}_{ab}$.
The Riemann sphere $\hat{\mathbb{C}}$ equipped with the random measures $\mu_\phi$ and $\nu_\phi$ is called the \emph{unit-area $\gamma$-quantum sphere}.

Next, we describe the random space-filling curve $\eta_{\hat{g}}$ arising from SLE$_{\kappa'}$ with $\kappa' = 16 / \gamma^2 \in (4,\infty)$.
A concise way to introduce this curve is via \emph{imaginary geometry}.
If $h : \mathbb{C} \to \mathbb{R}$ is a smooth real function, one can consider the flow lines of the complex vector field $e^{i h / \chi}$ where $\chi = \frac{\sqrt{\kappa'}}{2} - \frac{2}{\sqrt{\kappa'}}$. 
More precisely, to $z\in \mathbb{C}$ we associate the curve $\eta_z$ determined by
\begin{align}
  \eta_z'(t) = e^{\frac{i}{\chi}h(\eta_z(t))}, \quad \eta_z(0) = z, \quad t\geq 0.
\end{align}
Remarkably these flow lines are still well-defined if we take $h$ to be the (far from smooth) whole-plane GFF $h$ on $\hat{\mathbb{C}}$ (independently but similar to the one for LQG).
It can be shown, see \cite{Miller_Imaginary_2017}, that for each $z\in \mathbb{C}$ the flow-line $\eta_z(t)$ does not self-intersect and approaches $\infty$ as $t\to\infty$, and that for two distinct starting points $z,z' \in \mathbb{C}$ the flow lines $\eta_z$ and $\eta_{z'}$ almost surely eventually meet and stay together before reaching $\infty$ (Figure~\ref{fig:lqg-sle}b).
One may use this to associate an order to the points in the complex plane: $z$ precedes $z'$ if $\eta_z$ meets $\eta_{z'}$ from the left.
The space-filling curve SLE$_{\kappa'}$ is the continuous non-self-crossing path $\eta_{\hat{g}} : [0,1] \to \hat{\mathbb{C}}$ starting and ending at $\infty$ that visits the points of $\mathbb{C}$ in this order, parametrized such that $\eta_{\hat{g}}([s,t])$ has area $t-s$ for $0\leq s<t\leq 1$. 

Liouville Quantum Gravity for $\gamma \in (0,2)$ and the space-filling SLE$_{\kappa'}$ for $\kappa' \in (4,\infty)$ are intimately related to each other when $\kappa' = 16/\gamma^2$.
To formulate this let us consider a unit-area $\gamma$-quantum sphere $\mu_\phi$, $\nu_\phi$ and an independently-sampled space-filling curve $\eta_{\hat{g}}$.
It is then natural to consider the reparametrization $\eta_g$ of $\eta_{\hat{g}}$ such that it explores the quantum area at unit rate, meaning that $\mu_\phi(\eta_g([s,t])) = t-s$ for $0<s<t<0$.
Now for every $t \in (0,1)$, the traces $\eta_g([0,t])$ and $\eta_g([t,\infty])$ are closed subset of $\hat{\mathbb{C}}$. 
The boundary $\eta_g([0,t]) \cap \eta_g([t,\infty])$ at which they meet consists of two continuous curves starting at $\eta_g(t)$ and ending at $\infty$. 
Let $Z(t) = (L(t),R(t)) \in \mathbb{R}_{>0}^2$ be the $\nu_\phi$-lengths of these curves (Figure~\ref{fig:lqg-sle}d).
The crucial insight of the mating of trees approach \cite{duplantier2014liouville} is that the process $Z(t)$ has a very simple law.
To be precise, according to \cite[Theorem 1.1]{Miller_Liouville_2019} (and \cite[Theorem 1.3]{Ang_FZZ_2021} for the precise normalization) it has the law of a two-dimensional Brownian motion $(L(t),R(t))$ started from $(0,0)$ with covariance
\begin{equation}\label{eq:lqgcov}
    \mathrm{Var}(L(t))=\mathrm{Var}(R(t))=\frac{2}{\sin\left(\frac{\pi\gamma^2}{4}\right)}|t|, \hspace{2mm} \mathrm{and} \hspace{2mm} \mathrm{Cov}(L(t),R(t))=-2\cot\left(\frac{\pi\gamma^2}{4}\right)|t|
\end{equation}
and conditioned to stay in $[0,\infty)^2$ and to return to $(0,0)$ at time $t=1$.
Moreover, both the quantum sphere and the space-filling curve are almost surely determined by this process.
This means that, at least in principle, one can reconstruct the measures $\mu_\phi$ and $\nu_\phi$ as well as the curve $\eta_{\hat{g}}$ simply by looking at the Brownian excursion $Z(t)$.
In the next subsection we will discuss an explicit procedure.

At this point one should recognize the analogy with the discrete mating-of-trees bijections that we described above.
The quantum surface is the continuum analogue of the random planar map, while the space-filling SLE$_{\kappa'}$ is the analogue of the exploration path determined by the statistical system living on the planar map.
In the discrete case the bijections with lattice walks show that the random discrete surface is completely determined by a corresponding random walk, which starts at the origin and is conditioned to stay in the positive quadrant $\mathbb{Z}_{\geq 0}^2$ before returning to the origin after $n$ steps.
Upon rescaling the walk by $1/\sqrt{n}$ and normalizing the time to run over $[0,1]$, its law converges in a probabilistic sense to that of the Brownian excursion $(L(t),R(t))$.
Comparing \eqref{eq:covrho} to \eqref{eq:lqgcov} we observe that $\rho = - \cos (\pi \gamma^2/4)$ and therefore the string susceptibility $\gamma_{\mathrm{s}}$ must be related to $\gamma$ by 
\begin{equation}\label{eq:gammagammas}
  \gamma_{\mathrm{s}} = 1 - \frac{4}{\gamma^2},
\end{equation}
which is easily checked to be consistent with the relations between $\gamma_{\mathrm{s}}$, $\gamma$, $c$, $Q$ given in \eqref{gamma_string_with_c}, \eqref{eq:Q_gamma} and \eqref{conformal_charge}.

\begin{figure}
  \centering
  \includegraphics[width=.8\linewidth]{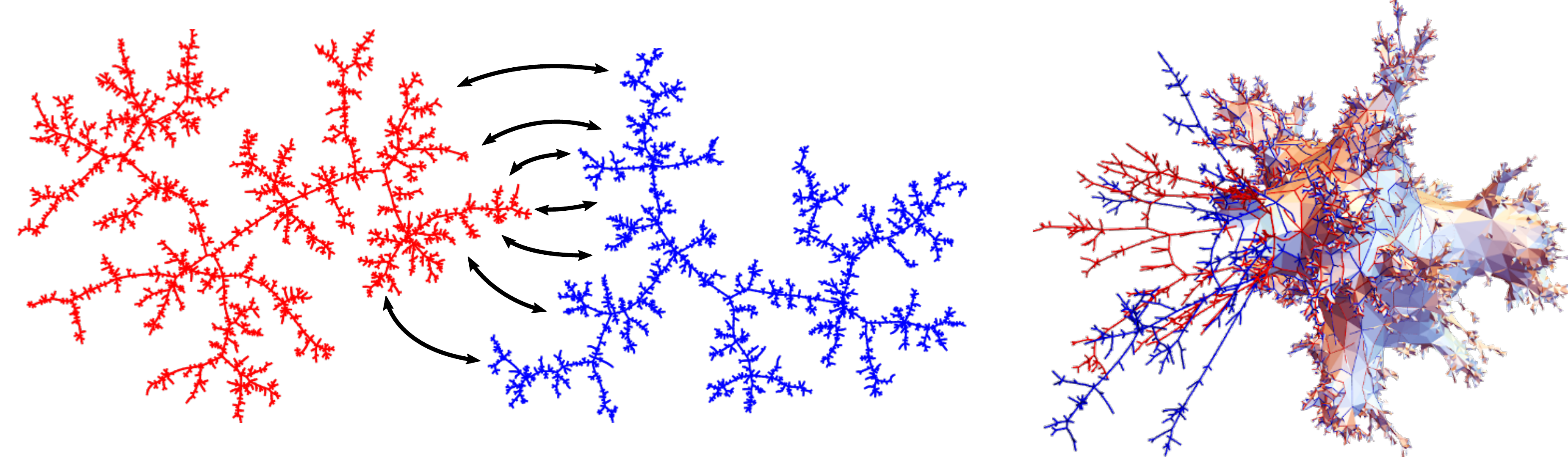}
  \caption{Illustration of the mating of trees construction. The excursions $L(t)$ and $R(t)$ in the positive half line each encode a real tree (in red and blue respectively). Upon pairwise identification of points in their contours a topological sphere emerges. The right figure illustrates an intermediate state in which only part of the contour is identified. \label{fig:treemating}}
\end{figure}

To wrap up, mating of trees provides a procedure to go back and forth between a unit-area $\gamma$-quantum sphere together with a space-filling SLE$_{\kappa'}$ curve on one side and a pair or correlated continuum random trees encoded by a Brownian excursion in the quadrant on the other, and each side (almost surely) determines the other. 

\subsection{Mated-CRT maps}\label{subsection:matedcrt}

As should be clear from Figure~\ref{fig:lqg-sle}b, the union of the flow lines $\eta_{z_1}, \ldots, \eta_{z_k}$ of points $z_1, \ldots, z_k\in\mathbb{C}$ has the structure of a tree spanning $z_1, \ldots, z_k$ and $\infty$.
If we increase the number $k$ of points this tree approaches a tree that spans the whole sphere, and the space-filling curve $\eta_{g}$ can be understood as tracing the contour of this tree.
The quantum length measure $\nu_\phi$ assigns a metric structure to the tree, and one can interpret the process $R_t$ as the distance in the tree between $\eta_g(t)$ and $\infty$.
Similarly, $L_t$ is the distance to $\infty$ along a complementary tree, which informally one can think of as what is left of the surface after the first tree is removed.  

The reconstruction of the quantum sphere from the two-dimensional Brownian excursion can also be understood from the perspective of the trees.
Both coordinates $L(t)$ and $R(t)$ describe a continuous excursion in the positive real line starting and ending at $0$.
Any such excursions $X : [0,1] \to \mathbb{R}_{\geq 0}$ naturally gives rise to a continuous metric space: the \emph{real tree} given by the unit interval $[0,1]$ with metric 
\begin{equation}
    d(s,t) = X(s) + X(t) - 2 \inf_{u\in [s,t]}X(u),
\end{equation}
where it is understood that we identify $s$ and $t$ whenever $d(s,t)=0$.
If $L(t)$ and $R(t)$ are uncorrelated, which happens for $\gamma = \sqrt{2}$, each is an independent Brownian excursion on the line and the corresponding real tree is called the Continuum Random Tree (CRT).
In general they encode a pair of correlated random trees very similar to the CRT.
It is relatively straightforward (see \cite[Section 1.3]{duplantier2014liouville}) to see that pairwise identification of points in the contours of the two trees leads to a space that is (almost surely) topologically equivalent to the 2-sphere. 
See Figure~\ref{fig:treemating} for an illustration.
What is not at all obvious is that the result has a natural conformal structure, let alone a natural metric.
A convenient way to see that it does is by considering successively finer discretizations of the surface as follows.

Consider an excursion $X : [0,1] \to \mathbb{R}_{\geq 0}$ in the positive real line such that $X(0)=X(1)=0$.
For any positive integer $n$ one may associate to $X$ a triangulation of the $n$-sided polygon with vertices labeled from $1$ to $n$ as follows \cite{gwynne2021tutte}.
We divide the interval $[0,1]$ into $n$ equal parts $[0,\tfrac1n],[\tfrac1n, \tfrac2n], \ldots,[1-\tfrac1n,1]$, one for each vertex.
For any $1 \leq x < y \leq n$ such that the vertices with labels $x$ and $y$ are not neighbours, we draw a diagonal connecting these vertices if there is a horizontal segment below the graph of $X$ connecting the intervals $[\frac{x-1}{n},\frac{x}{n}]$ and $[\frac{y-1}{n},\frac{y}{n}]$ in the graph, i.e.\ if there is an $s \in [\frac{x-1}{n},\frac{x}{n}]$ and a $t \in [\frac{y-1}{n},\frac{y}{n}]$ such that $X(s) = X(t)$ and $X(u) \geq X(s)$ for all $u\in [s,t]$.
For generic $X$, for instance when $X$ is a Brownian excursion, the result is a triangulation.

\begin{figure}[H]
  \centering
    \includegraphics[width=1.0\textwidth]{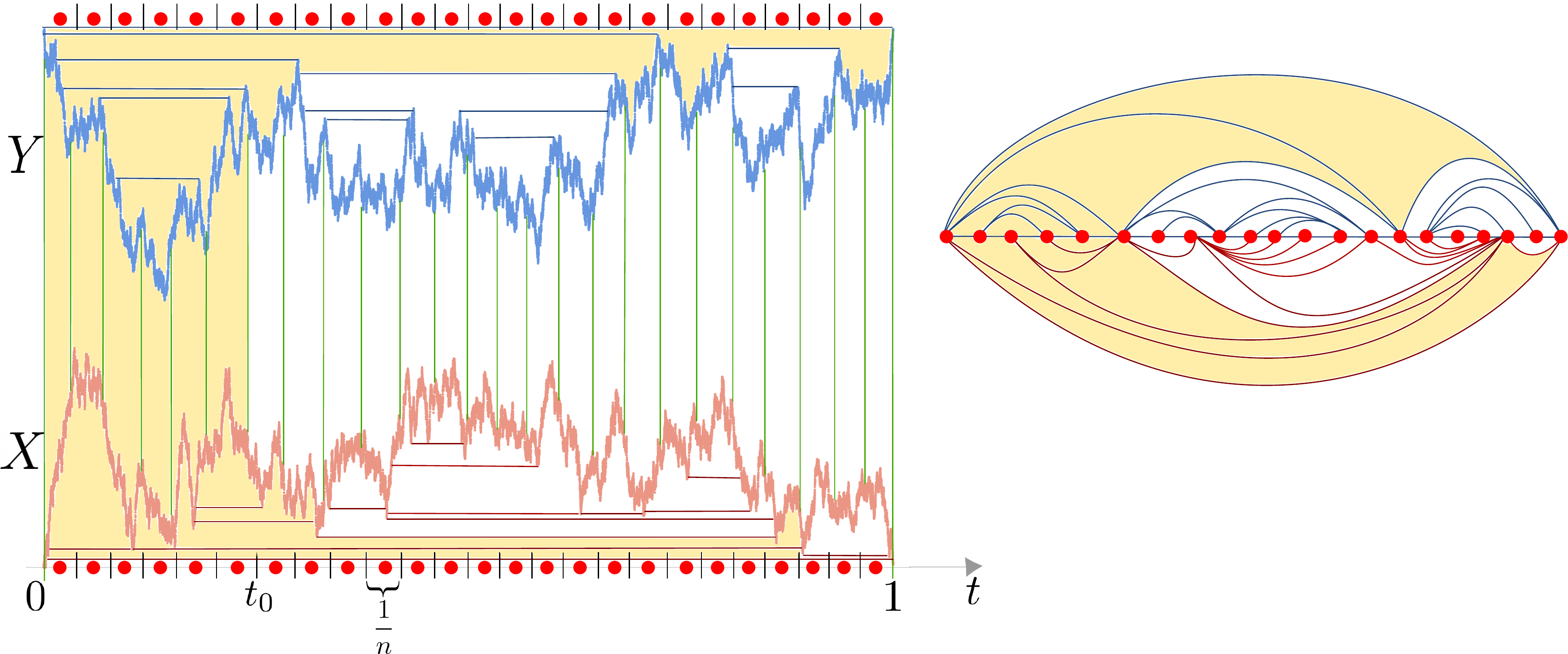}
  \caption{Left: The components $(X,Y)$ of a 2D Brownian excursion are drawn with $Y$ drawn upside down for illustration purposes. The interval $[0,1]$ is divided in $n$ equal parts and each of them correspond to a vertex. The horizontal segments that lead to edges between vertices are indicated. Right: The triangulation resulting from gluing the pair of triangulated $n$-gons (the one associated to $Y$ in blue on top and the one associated to $X$ on the bottom).
  Note that the top and bottom arc are yet to be identified.
  The shaded region represents the region explored by the space-filling curve up to time $t_0$.}
  \label{fig:crtmating}
\end{figure}

In the case of a two-dimensional Brownian excursion $(L(t),R(t))$ we thus naturally obtain a pair of triangulated polygons by applying the construction to both coordinates.
Gluing these two polygons together produces a triangulation of the 2-sphere with $2n-4$ triangles, called the \emph{Mated-CRT map} \cite{gwynne2021tutte}.
It is naturally equipped with a Hamiltonian cycle, i.e.\ a simple closed path on the triangulation that visits all vertices (in this case in order of their labels $1,\ldots,n$), and rooted on the edge that connects the vertices with label $1$ and $n$.
See Figure~\ref{fig:crtmating} for an example.
This random triangulation is very much analogous to the discrete models in sections \ref{sec:stquad} and \ref{sec:perctri}, except that it exists for any of the universality classes parametrized by $\gamma\in(0,2)$.
For any $n$ we obtain a unit-area Riemannian metric $g^{(n)}_{ab}$ on the 2-sphere by interpreting each triangle as an equilateral Euclidean triangle of area $1/(2n-4)$.
Informally, the metric associated to the mating of trees should be obtained as the large-$n$ limit of these metrics $g^{(n)}_{ab}$.
More precisely, one can work with the canonical Tutte embedding of the triangulation in the sphere and it is shown in \cite[Theorem 1.1 and Remark 3.7]{gwynne2021tutte} that the corresponding measure $\sqrt{g^{(n)}}\mathrm{d}^2 z$ converges (in a weak sense) to $\mu_\phi$ as $n\to\infty$ and the Hamiltonian cycle to the space-filling curve $\eta_{g}$.

So far we have not discussed geodesic distances in the quantum sphere.
Naively, one would expect to find a metric structure, i.e.\ a distance between $x,y\in \hat{\mathbb{C}}$, by minimizing the quantum length $\nu_\phi(\Gamma([0,1]))$ of a curve $\Gamma$ from $x$ to $y$.
But due to the fractal nature of the geometry, this limit is identically zero. 
Instead one should consider a different regularization \cite{Gwynne2021}, namely there exists a deterministic positive real number $d_\gamma > 2$ depending only on $\gamma$ such that the regularized distance 
\begin{align}
  D_\epsilon(x,y) = \frac{\inf_\Gamma \int_{\Gamma} e^{\frac{\gamma}{d_\gamma} \phi_\epsilon(z)}(\hat{g})^{\frac{1}{2d_\gamma}}|\mathrm{d}z|}{\left(\int_{\hat{\mathbb{C}}} e^{\gamma \phi_\epsilon(z)}\sqrt{\hat{g}}\,\mathrm{d}^2z\right)^{1/d_\gamma}}
\end{align}
possesses a well-defined limiting metric $D_\phi(x,y)$ (in probability) as $\epsilon\to 0$ when appropriately rescaled (by a factor of order $\epsilon^{-1 + \frac{2}{d_{\gamma}}}$ in $\epsilon$).
The value $d_\gamma$ is precisely the \emph{Hausdorff dimension} of this metric \cite{Gwynne2019}, which informally is saying that the $\mu_\phi$-quantum area of a geodesic ball of radius $r$ around any point is of order $r^{d_\gamma}$ when $r \to 0$.
The exact value of $d_\gamma$ is only known for $\gamma = \sqrt{8/3}$, corresponding to the pure gravity universality class, where $d_{\sqrt{8/3}} = 4$.
For $\gamma \neq \sqrt{8/3}$, rigorous bounds are known \cite{Ding:2018uez,Ang2019} as well as numerical estimates \cite{Barkley_2019}. Moreover, as $\gamma\rightarrow 0$ the dimension $d_\gamma$ approaches $2$ (see \cite{ding2019upper} for bounds on the convergence rate) in accordance with the constant curvature solution $g_{ab}$ to the classical Liouville action at $\gamma=0$.

It is widely expected that this random metric $D_\phi(x,y)$ agrees with the large-$n$ limit of the graph distance within the $n$-vertex Mated-CRT map when normalized by $n^{-1/d_\gamma}$, and also with the geodesic distance as measured by $g_{ab}^{(n)}$ with the same normalization.
A proof is still out of reach, but it is known \cite[Theorem 1.6]{Ding:2018uez} that the number of vertices in a ball of radius $r$  around a randomly chosen vertex in the limit $n\to\infty$ grows like $r^{d_\gamma}$ with increasing radius $r$.
Therefore the simple model of Mated-CRT maps can be used to estimate the Hausdorff dimension of Liouville Quantum Gravity for any $\gamma \in (0,2)$. 
We will pursue this avenue in the next section.

The string susceptibility $\gamma_{\mathrm{s}}$ can also be interpreted at the level of the Mated-CRT maps in terms of the distribution of sizes of \emph{minimal-neck baby universes} (minbus) within the geometry.
Since the Mated-CRT map is a loopless triangulation, the minimal length of a simple closed cycle is two. 
We let a \emph{minbu of size $k$} for $2 \leq k \leq n-2$ be a connected region of $2k-2$ triangles not containing the root edge that is separated by a cycle of length two from the remaining $2(n-k)-2$ triangles.
The string susceptibility is often introduced \cite{Jain:1992bs} as the exponent featuring in the expected number $E_{n,k}$ of minbus of size $k$,
\begin{equation}\label{eq:minbuscaling}
    \lim_{n\to\infty} \frac{E_{n,k}}{n} = C\, k^{\gamma_{\mathrm{s}} - 2} + o(k^{\gamma_{\mathrm{s}}-2})\qquad \text{as }k\to\infty.
\end{equation}
Let us verify that this definition agrees with the relation \eqref{eq:gammagammas}.

Note from the construction of the Mated-CRT map that a minbu of size $k$ is associated to any $x=1,\ldots,n-k$ for which both triangulated polygons have a diagonal connecting $x$ to $x+k$.
Therefore $E_{n,k} / n$ is the probability of this event when $x$ is sampled uniformly.
Denoting $X(t) = (X_1(t),X_2(t)) = (L(t),R(t))$, this happens precisely when 
\begin{equation*}
    \min_{u \in [\frac{x}{n},\frac{x+k-1}{n}]} X_i(u) > \max\left( \min_{s\in [\frac{x-1}{n},\frac{x}{n}]}X_i(s),\min_{t\in [\frac{x+k-1}{n},\frac{x+k}{n}]}X_i(t)\right)\qquad \text{for }i=1,2.
\end{equation*}
In the limit $n \to \infty$ the probability is the same as that for an unrestricted correlated two-dimensional Brownian motion $(\tilde{X}_1(t),\tilde{X}_2(t))$, such that
\begin{equation}
    \lim_{n\to\infty} \frac{E_{n,k}}{n} = \mathbb{P}\left(\min_{u \in [x,x+k-1]} \tilde{X}_i(u) > \max\left( \min_{s\in [x-1,x]}\tilde{X}_i(s),\min_{t\in [x+k-1,x+k]}\tilde{X}_i(t)\right)\text{ for }i=1,2\right).
\end{equation}
But this is essentially the probability that a two-dimensional correlated Brownian motion started close to the origin, remains in the quadrant for time at least $k$ and is close to the origin again at time $k$.
This can be estimated using the heat kernel of the Brownian motion \cite[Lemma 1]{Banuelos1997}, and scales with $k$ as $k^{-1-\gamma^2/4}$ (see discussion about Brownian motion in the wedge below).
We thus find 
\begin{equation}
    \lim_{n\to\infty} \frac{E_{n,k}}{n} = C\, k^{-1-\gamma^2/4} + o(k^{-1-\gamma^2/4})\qquad \text{as }k\to\infty.
\end{equation}
This is clearly in agreement with \eqref{eq:gammagammas} and \eqref{eq:minbuscaling}.

\section{Mated-CRT graphs from multi-dimensional Brownian excursions}\label{sec:matedcrtgeneralized}

\subsection{Mated-CRT graphs}\label{sec:matedcrtgraphs}

\begin{figure}[H]
  \centering
    \includegraphics[width=0.8\textwidth]{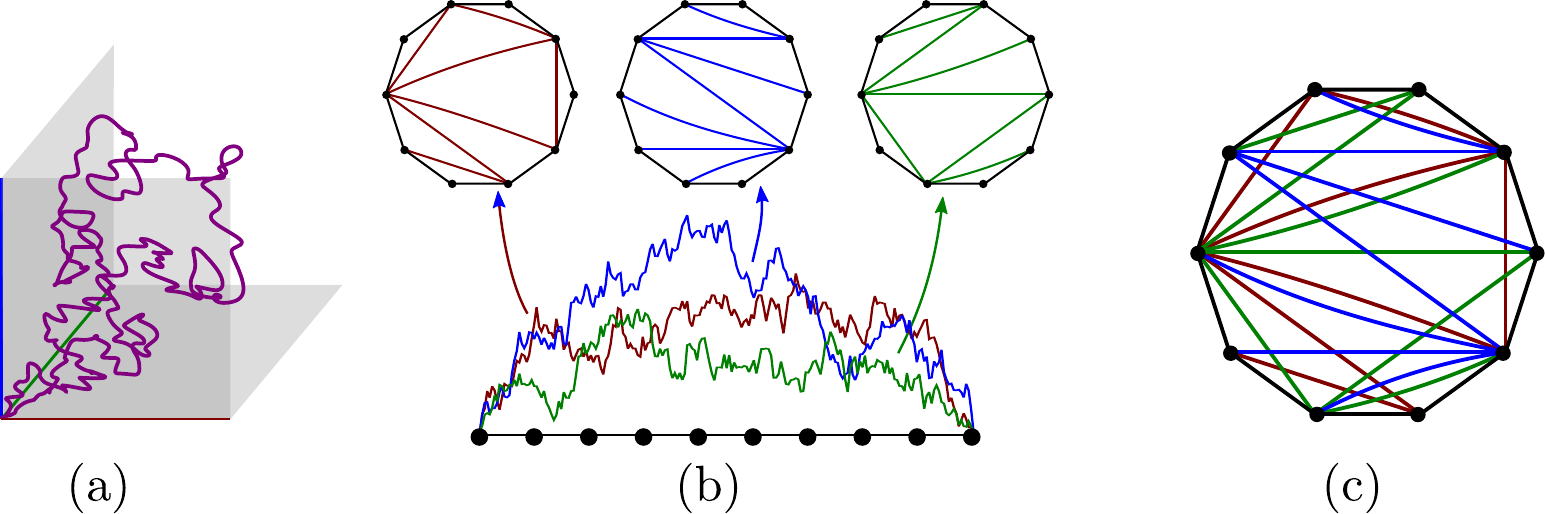}
  \caption{(a) Illustration of a three-dimensional Brownian excursion. (b) To each of the $d=3$ components of the excursion we may associate a triangulation of the $n$-gon. (c) The resulting Mated-CRT graph $G^{\mathbf{C}}_n$ with the Hamiltonian cycle appearing in black. 
  \label{fig:generalizedmodel}}
\end{figure}

Now that we know how to read metric properties and the string susceptibility from the combinatorial data of a Mated-CRT map, let us introduce a natural generalization.
Let $d = 2,3,\ldots$ and $\mathbf{C}$ be a real positive-definite symmetric $d\times d$ matrix. 
Then we may consider $d$-dimensional Brownian motion $X(t) = (X_1(t),\ldots, X_d(t))$ started at the origin in $\mathbb{R}^d$ with covariance matrix $\operatorname{Cov}(X_i(t),X_j(t)) = \mathbf{C}_{ij} |t|$.
A \emph{Brownian excursion} with covariance $\mathbf{C}$ is then such a Brownian motion for $t\in [0,1]$ that is conditioned to start and end at the origin and stay in the octant $\mathbb{R}_{>0}^d$ for $t\in (0,1)$.
We can associate to this Brownian excursion a random (multi-)graph $G^{\mathbf{C}}_n$ on $n$ vertices with a distinguished Hamiltonian cycle by gluing the $d$ triangulated $n$-gons associated to the $d$ excursions $X_1(t),\ldots, X_d(t)$ along their boundary (Figure~\ref{fig:generalizedmodel}).
For $d=2$ this graph is planar and $G^{\mathbf{C}}_n$ corresponds to the graph underlying the Mater-CRT map, while for $d \geq 3$ the graph is generally non-planar.

The central question is whether the graph $G^{\mathbf{C}}_n$, seen as a metric space induced by the graph distance, possesses a scaling limit, meaning that there exists some real number $d_H^{\mathbf{C}} > 0$ for which the rescaled metric space $n^{-1/d_H^{\mathbf{C}}}G^{\mathbf{C}}_n$ has a continuous limit as $n\to\infty$ (in a Gromov--Hausdorff sense).
A positive answer for $d\geq 3$ would give rise to new families of universality classes of random geometries, which based on the two-dimensional case one would expect to depend on the covariance matrix $\mathbf{C}$. 
Note that the construction of $G^{\mathbf{C}}_n$ is invariant under scaling of the coordinate axes, and its law therefore is invariant under coordinate-wise rescaling of the matrix $\mathbf{C}$.
We may thus assume unit diagonal entries of $\mathbf{C}$ without loss of generality, and we are left with a $d(d-1)/2$-dimensional phase space of models.
In $d=2$ this 1-dimensional phase space is parametrized by the Liouville coupling $\gamma \in (0,2)$.

As a first indication of a non-trivial scaling limit for $d\geq 2$, we will compute the string susceptibility of $G^{\mathbf{C}}_n$. 
The definition of a minbu (minimal-neck baby universe) is easily extended to $G^{\mathbf{C}}_n$: a minbu of size $k$ in $G^{\mathbf{C}}_n$ is a pair of vertices with label $x$ and $x+k$, such that removing $x$ and $x+k$ and all incident edges from $G^{\mathbf{C}}_n$ one is left with two connected components with $n-k-1$ and $k-1$ vertices respectively.
Following the same reasoning as in the two-dimensional case the expected number $E_{n,k}^{\mathbf{C}}$ of minbus of size $k$ satisfies
\begin{equation*}
    \lim_{n\to\infty} \frac{E_{n,k}^{\mathbf{C}}}{n} = C k^{\gamma_s-2} + o(k^{\gamma_s-2}),
\end{equation*}
if the heat kernel $P^{\mathbf{C}}_t(x,y)$ of the $d$-dimensional Brownian motion with covariance matrix $\mathbf{C}$ on the octant with absorbing boundary conditions falls off like $P^{\mathbf{C}}_t(x,y) = c\, t^{\gamma_s-2} + o(t^{\gamma_s-2})$ as $t\to\infty$.
Let us take a closer look at this process to see whether this is realized.

\subsection{Brownian Excursions in a cone}

Instead of dealing with correlated Brownian motion in the octant $\mathbb{R}_{\geq 0}^d$, it is often more convenient to work with uncorrelated Brownian motion in an appropriate cone $W \subset \mathbb{R}^d$. 
If $\mathbf{C}$ is a positive-definite symmetric matrix, then we can find an invertible real $d\times d$ matrix $R$ such that $\mathbf{C} = R R^T$ (in fact $R$ may be taken to be lower-triangular with positive entries on the diagonal, in which case it is called the Cholesky decomposition of $\mathbf{C}$).
Let $W = R^{-1}\mathbb{R}_{\geq 0}^d$ be the preimage of the octant by the linear map $R$.
Then the standard $d$-dimensional Brownian motion in the cone $W$ is mapped by $R$ to a Brownian motion with covariance matrix $\mathbf{C}$ in the octant. 

The corresponding heat kernel $P^{\mathbf{C}}_t(x,y) \mathrm{d}^dy$ measures the probability density that a standard Brownian motion started at $x \in W$ remains within $W$ for at least time $t$ and is located at $y\in W$ at time $t$. 
By separation of radial and angular motion, it can be explicitly expressed in terms of the orthonormal eigenmodes of the spherical Laplace-Beltrami operator $L_{\mathbb{S}^{d-1}}$ on the spherical region $W \cap \mathbb{S}^{d-1} \subset \mathbb{R}^d$ with Dirichlet boundary conditions,
\begin{equation}\label{angular_dirichlet}
    \begin{cases}
        L_{\mathbb{S}^{d-1}}m_i(\tilde{x}) = -\lambda_im_i(\tilde{x}) \hspace{2mm} & \mathrm{for} \hspace{2mm}  \tilde{x}\in W\cap\mathbb{S}^{d-1},\\
        m_i(\tilde{x})  =0 \hspace{2mm} & \mathrm{for} \hspace{2mm}  \tilde{x}\in \partial W\cap\mathbb{S}^{d-1}.
    \end{cases}
\end{equation}
Namely \cite[Lemma 1]{Banuelos1997}
\begin{equation}\label{eq:heatkernel}
    P^{\mathbf{C}}_t(x,y) = \frac{e^{-\frac{|x|^2+|y|^2}{2t}}}{|x|^{\frac{d}{2}-1}|y|^{\frac{d}{2}-1}} \sum_{j=1}^\infty \frac{1}{t} I_{\alpha_j}\left(\frac{|x||y|}{t}\right) m_j\left(\frac{x}{|x|}\right)m_j\left(\frac{y}{|y|}\right),
\end{equation}
where $I_{\alpha}(r)$ is a modified Bessel function satisfying $I_{\alpha}(r) \sim r^{\alpha}$ as $r\to 0$ and 
\begin{equation}
    \alpha_j = \sqrt{ \lambda_j + \left(\frac{d}{2}-1\right)^2}.
\end{equation}
It follows that for fixed $x,y\in W$,
\begin{equation}\label{eq:heatkernelasymp}
    P^{\mathbf{C}}_t(x,y) = c\, t^{-\alpha_1-1} + o(t^{-\alpha_1-1})
\end{equation}
as $t\to\infty$, where the exponent depends on the fundamental eigenvalue $\lambda_1$ of $L_{\mathbb{S}^{d-1}}$ on $W \cap \mathbb{S}^{d-1}$.
Hence the string susceptibility of the mated-CRT graph with covariance matrix $\mathbf{C}$ is \begin{equation}
    \gamma_{\mathrm{s}} = 1 - \alpha_1 = 1 - \sqrt{\lambda_1 + \left(\frac{d}{2}-1\right)^2}.\label{string_sus}
\end{equation}

\begin{figure}[H]
  \centering
    \includegraphics[width=0.8\textwidth]{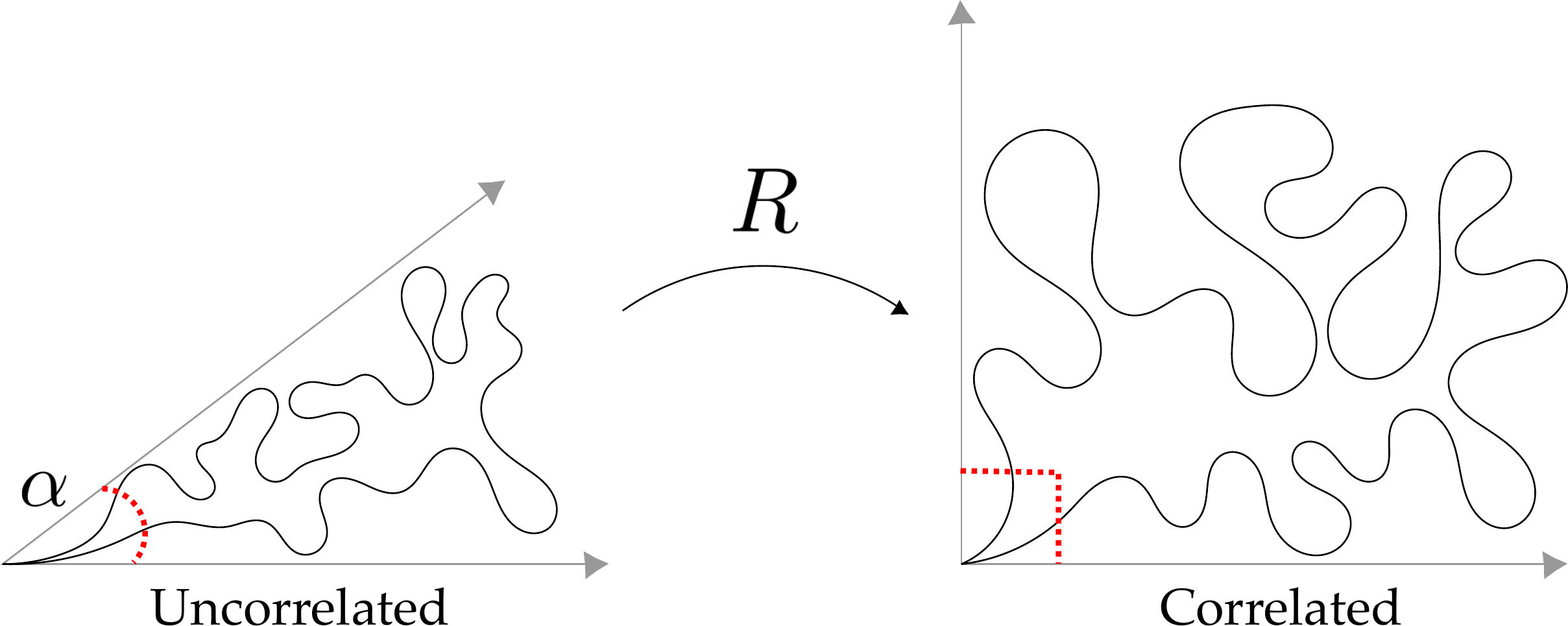}
  \caption{An uncorrelated BE in a cone of opening angle $\alpha$ is mapped to a Correlated BE in $\mathbb{R}_{\geq 0}^2$ by \eqref{2D_rotation}.}
  \label{fig:2D_wedge}
\end{figure}

In the two-dimensional case, the appropriate linear transformation $R$ is 
\begin{equation}
R=\begin{pmatrix}
 \sin \alpha & -\cos \alpha \\
 0 & 1 
\end{pmatrix}, \qquad \mathbf{C} = R R^T = \begin{pmatrix}
1& -\cos \alpha \\
-\cos \alpha & 1 
\end{pmatrix}.\label{2D_rotation}
\end{equation}
Then $R^{-1}\mathbb{R}_{\geq 0}^2$ is a cone of opening angle $\alpha$, with the right boundary ray along the positive x-axis (Figure~\ref{fig:2D_wedge}).
The corresponding fundamental eigenmode is $m_1(\theta) = \sin( \pi \theta / \alpha)$ with eigenvalue $\lambda_1 = \pi^2 / \alpha^2$.
We see that $\gamma_{\mathrm{s}} = 1 - \pi / \alpha = 1 - 4 / \gamma^2$ is consistent with \eqref{eq:gammagammas}.

\begin{figure}[H]
  \centering
    \includegraphics[width=0.3\textwidth]{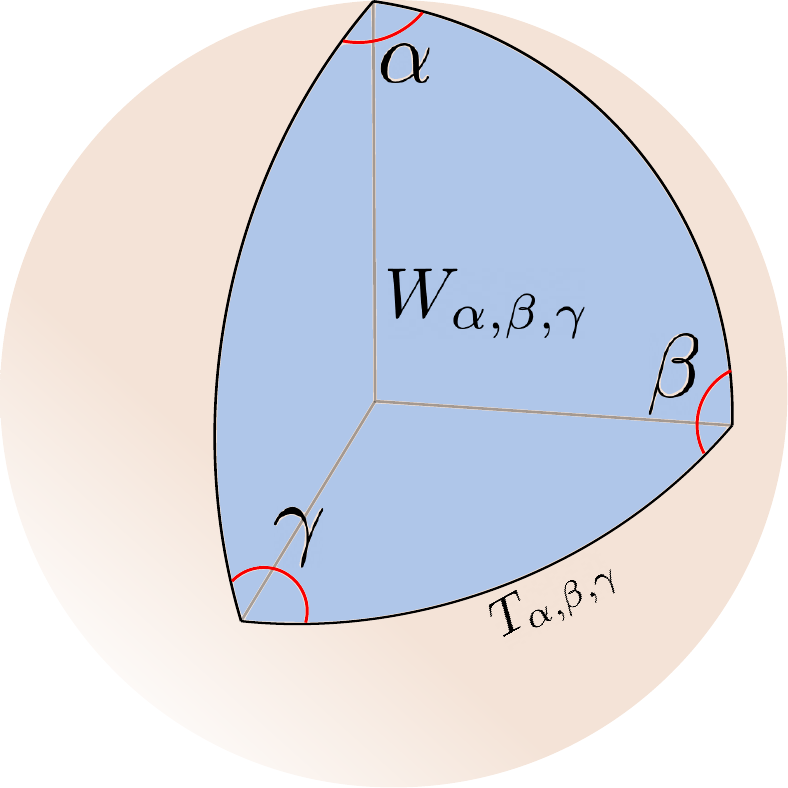}
  \caption{The cone $W_{\alpha,\beta,\gamma}$ corresponds to the solid spherical region delimited by the spherical triangle $T_{\alpha,\beta,\gamma}$ in the unit sphere.}
  \label{fig:3D_wedge}
\end{figure}

\begin{figure}[H]
  \centering
    \includegraphics[height=6cm]{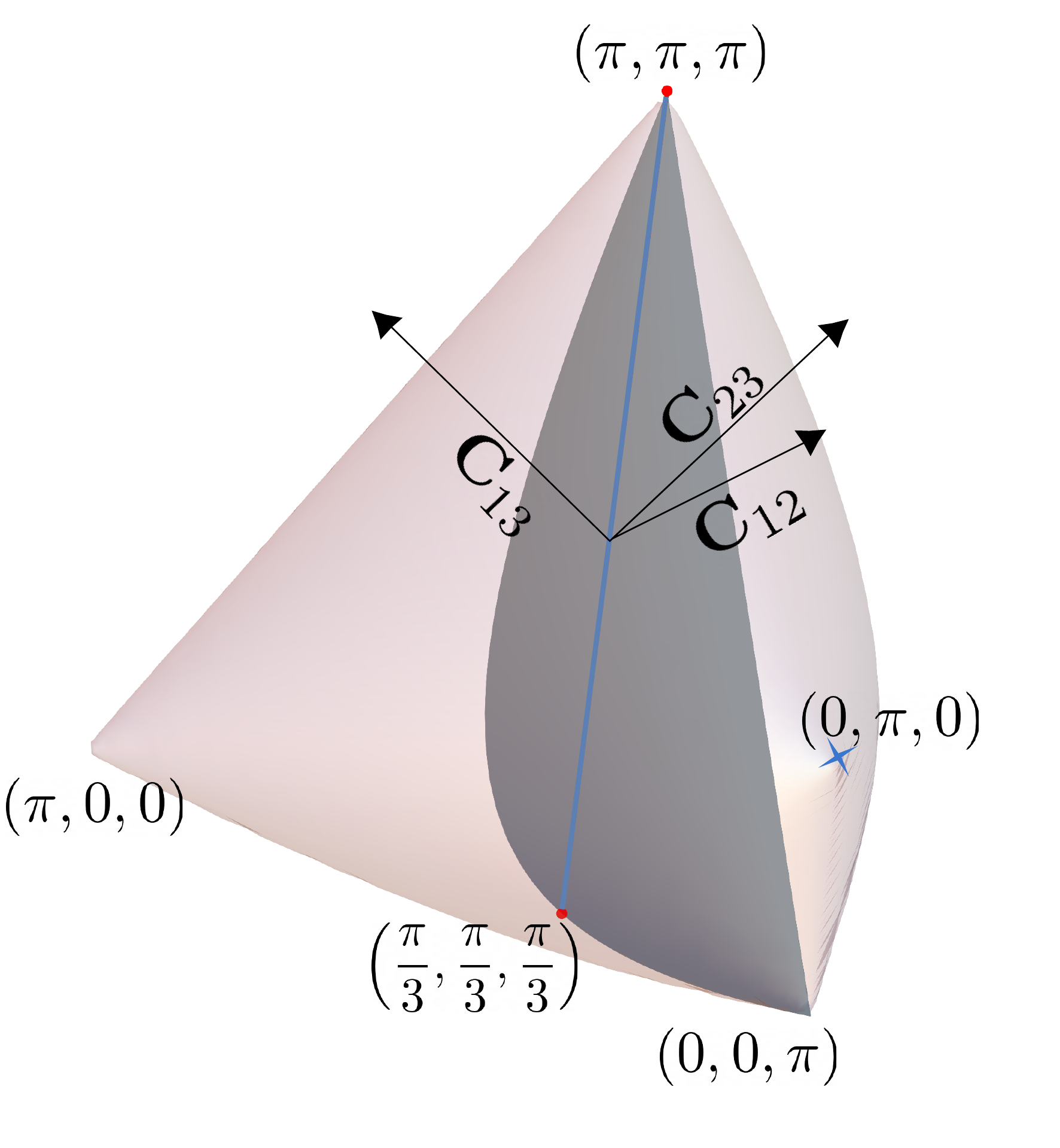}%
    \includegraphics[height=6cm]{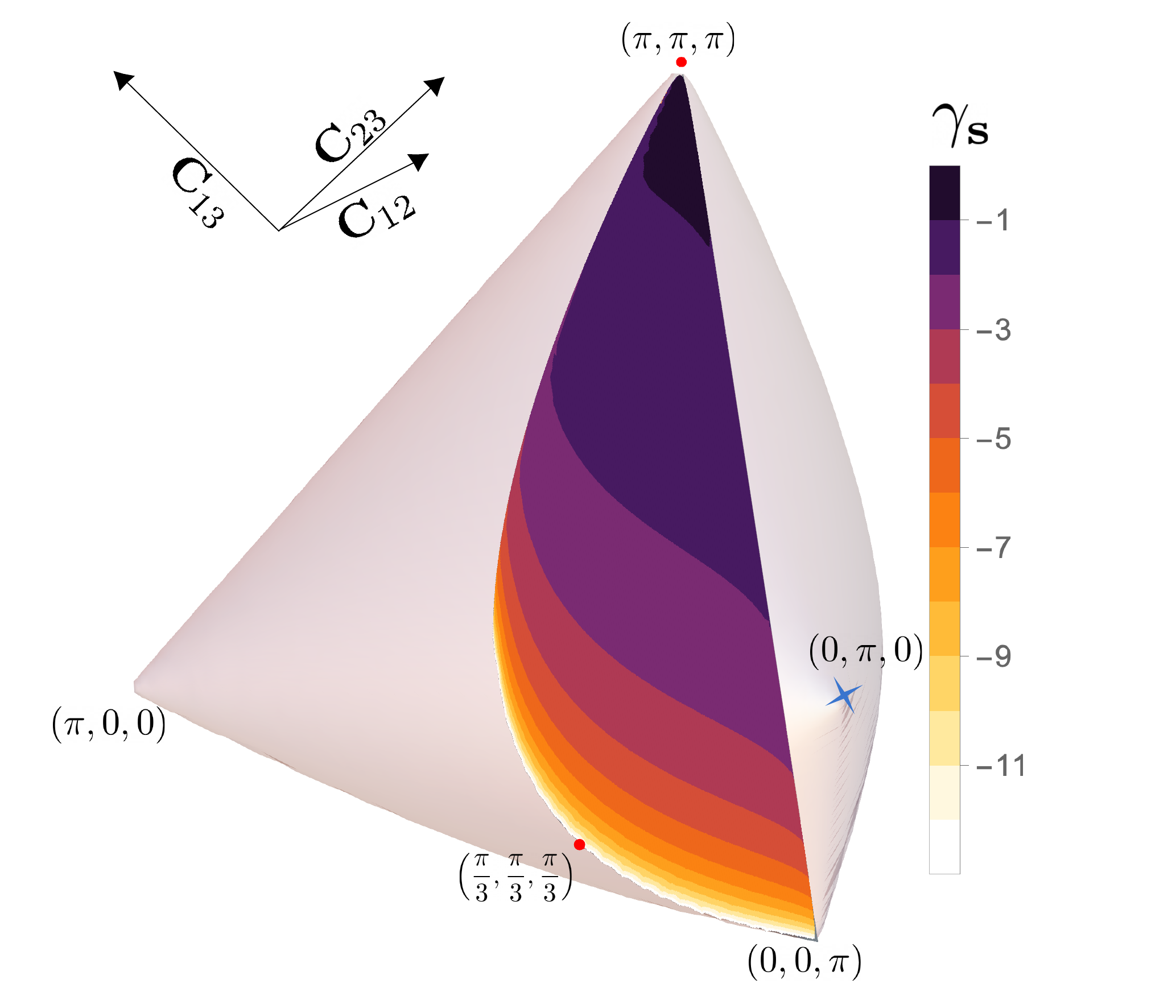}
  \caption{Left: The phase space of covariance matrices $\mathbf{C}$, parametrized by $\mathbf{C}_{12}=-\cos{(\alpha)}$, $\mathbf{C}_{23}=-\cos{(\beta)}$, $\mathbf{C}_{13}=-\cos{(\gamma)}$, spans the ``tetrahedral'' region $\mathbf{C}_{12}^2+\mathbf{C}_{13}^2+\mathbf{C}_{23}^2 - 2 \mathbf{C}_{12}\mathbf{C}_{13}\mathbf{C}_{23} < 1$. The corners are labeled with their spherical angles $(\alpha,\beta,\gamma)$. The blue diagonal corresponds to  equilateral spherical triangles ($\alpha=\beta=\gamma$), while the gray plane indicates the isosceles spherical triangles ($\beta=\gamma$). The red dots are the smallest ($\frac{\pi}{3},\frac{\pi}{3},\frac{\pi}{3}$) and largest ($\pi,\pi,\pi$) equilateral triangles, which will be of particular interest in the results. Right: estimates of the string susceptibility $\gamma_{\mathrm{s}}$ for isosceles triangles obtained from Finite Element Methods.}
  \label{fig:cov_tetrahedron}
\end{figure}

In the three-dimensional case the most general positive-definite symmetric matrix $\mathbf{C}$ with unit diagonal entries is given by \cite{Bogosel20203DPL}
\begin{equation}
\mathbf{C}=\begin{pmatrix}
    1 & -\cos(\alpha) & -\cos(\gamma) \\
    -\cos(\alpha) & 1 & -\cos(\beta) \\
    -\cos(\gamma) & -\cos(\beta) & 1
\end{pmatrix}\qquad \text{with }\alpha,\beta,\gamma \in (0,\pi), \, \alpha + \beta + \gamma > \pi.\label{cov3d_general}
\end{equation}
The corresponding cone $W_{\alpha,\beta,\gamma}=R^{-1}\mathbb{R}_{\geq 0}^3$ intersects $\mathbb{S}^2$ in a spherical triangle $T_{\alpha,\beta,\gamma}$ with angles $\alpha$, $\beta$ and $\gamma$ (Figure~\ref{fig:3D_wedge}) and its corresponding 3-dimensional phase space is the interior of a rounded tetrahedron (Figure~\ref{fig:cov_tetrahedron}). 
The eigenvalue $\lambda_1$ of the fundamental eigenmode $m_1$ of $T_{\alpha,\beta,\gamma}$ is only known for special values of the angles. For example, for birectangular spherical triangles ($\beta=\gamma=\pi/2$) the fundamental eigenvalue is known to be \cite{eigenvalues_spher}  
\begin{equation}
    \lambda_1=\left(1 +\frac{\pi}{\alpha}\right)\left(2 +\frac{\pi}{\alpha}\right),
\end{equation}
corresponding to a string susceptibility of $\gamma_{\mathrm{s}} = -\frac{1}{2}-\frac{\pi}{\alpha}$. 
On the other hand, for spherical triangles with very small area $\alpha+\beta+\gamma-\pi$ the fundamental eigenvalue is well approximated by that of the Laplacian on a Euclidean triangle with angles $\alpha,\beta,\pi-\alpha-\beta$ and area $\alpha+\beta+\gamma-\pi$. 
Denoting the fundamental eigenvalue of the unit-area Euclidean triangle with these angles by $\lambda_{\text{Eucl}}(\alpha,\beta)$, we thus have
\begin{equation}
    \lambda_1 = \frac{\lambda_{\text{Eucl}}(\alpha,\beta)}{\alpha+\beta+\gamma-\pi} + O(1)\qquad\text{as }\gamma\to \pi-\alpha-\beta.
\end{equation}
In the case of the equilateral triangle, it is a classical computation that $\lambda_{\text{Eucl}}(\tfrac{\pi}{3},\tfrac{\pi}{3}) = \tfrac{4}{\sqrt{3}}\pi^2$.
Hence, for the equilateral case we find the string susceptibility
\begin{equation}
    \gamma_{\mathrm{s}} = -\frac{2\pi}{3^{3/4}\sqrt{\alpha-\tfrac\pi3}} + 1 + O\left(\sqrt{\alpha-\tfrac\pi3}\right).\qquad (\alpha = \beta = \gamma)
\end{equation}
We see that $\gamma_\mathrm{s}\rightarrow-\infty$ as $\alpha\rightarrow\frac{\pi}{3}$ analogous to the behaviour of the string susceptibility in 2D when the cone becomes very narrow. The maximum is reached near the maximal area region in both cases. For more general regions we use the Finite Element Method (FEM) to determine the solutions numerically\footnote{Recently, more precise methods to compute these fundamental eigenvalues have been developed in \cite{dahne2020computation}.}. See Figure~\ref{fig:cov_tetrahedron}.

\subsection{A simpler biased Brownian excursion}\label{sec:biasedbm}

Unit-time Brownian excursions in a non-trivial cone $W$ are challenging objects to simulate efficiently.
For this reason we introduce a slightly biased version of the Brownian excursion, which is easier to simulate.
As we will see in a minute, for a unit-time Brownian excursion the integral $\int_0^1 |X(t)|^{2\alpha_1-2} \mathrm{d} t$ has finite expectation value $C_{\alpha_1} = 2^{\alpha_1-1} \Gamma(\alpha_1) / \alpha_1 > 0$.
We may thus introduce the Brownian excursion $\hat{X}(t)$ obtained from $X(t)$ by biasing its law by the value of this integral, meaning that the probability measure of $\hat{X}(t)$ is that of $X(t)$ multiplied by $\int_0^1 |X(t)|^{2\alpha_1-2} \mathrm{d} t / C_{\alpha_1}$.
In probabilistic terms, the new random excursion $\hat{X}(t)$ is absolutely continuous with respect to the unit-time Brownian excursion and therefore displays the same critical exponents.
In particular, the Mated-CRT graph $\hat{G}^{\mathbf{C}}_n$ associated to the excursion $\hat{X}(t)$ will display the same local geometry as the unbiased Mated-CRT graph $G^{\mathbf{C}}_n$ when $n \to \infty$, and therefore agree on the Hausdorff dimension $d_\gamma$ (if it exists) and the string susceptibility $\gamma_s$.

This is quite useful, because we claim that $\hat{X}(t)$ can be more easily sampled than $X(t)$.
Let $S$ be a random point chosen from the spherical region $W \cap \mathbb{S}^{d-1}$ with probability distribution $m_1(x)^2$ (recall that this eigenmode is assumed to be normalized and thus $m_1(x)^2$ integrates to one on $W \cap \mathbb{S}^{d-1}$).
Independently, we sample two independent Brownian motions $X^1(t)$ and $X^2(t)$ started at $S$ and conditioned to touch the boundary $\partial W$ at the origin.
If $T_1$ and $T_2$ are the hitting times of the origin, then we make the identification 
\begin{equation}\label{eq:bmconcat}
    \hat{X}(t) = \frac{1}{\sqrt{T_1+T_2}}\begin{cases}
        X^1(T_1 - t (T_1+T_2)) & 0 \leq t \leq \frac{T_1}{T_1+T_2} \\
        X^2(t (T_1+T_2) - T_1) & 1 \geq t \geq \frac{T_1}{T_1+T_2}
    \end{cases}.
\end{equation}
In words, we concatenate the reversal of the first curve with the second to produce an excursion from the origin, which is then rescaled to have unit duration (taking into account the usual Brownian scaling relations).

To understand why this works, let us first compute the distribution of the hitting time $T_1$ (which is the same as that of $T_2$).
From the $y\to 0$ limit of the heat kernel \eqref{eq:heatkernel} with $|x|=1$,
\begin{equation*}
    P_t^{\mathbf{C}}(x,y) \stackrel{y\to 0}{\sim} t^{-\alpha_1-1} e^{-\frac{1}{2t}} |y|^{\alpha_1+1-\frac{d}{2}} m_1\left(\frac{y}{|y|}\right)m_1\left(x\right),\qquad (|x|=1)
\end{equation*}
it follows that the hitting time $T_1$ is independent of the starting position $S$ and distributed as an \emph{inverse gamma distribution} with index $\alpha_1$ and scale $1/2$, i.e.\ has density
\begin{equation}\label{eq:invgamma}
    \frac{1}{2^{\alpha_1} \Gamma(\alpha_1)}t^{-\alpha_1-1}e^{-\frac{1}{2t}} \mathrm{d}t\qquad \text{on }\mathbb{R}_{>0}.
\end{equation}

Next, we determine the density of the Brownian excursion $X(t)$ in $W$ at a fixed time $s\in(0,1)$, which is obtained from the heat kernel \eqref{eq:heatkernel} via the limit 
\begin{align}
    \lim_{y_1,y_2\to 0} \frac{P_{s}(y_1,x)P_{1-s}(x,y_2)}{P_{1}(y_1,y_2)} 
    &= \frac{1}{s^{\alpha_1+1}(1-s)^{\alpha_1+1}} |x|^{2\alpha_1+1} e^{-\frac{|x|^2}{2}\left( \frac{1}{s}+\frac{1}{1-s}\right)} |x|^{1-|d|} m_1\left(\frac{x}{|x|}\right)^2.
\end{align}
It follows that $X(s) / |X(s)|$ for any $s$ is distributed like the point $S$ above and that the distance $|X(s)|$ to the origin has probability density
\begin{align}\label{eq:excdistancedist}
    \rho_{s}(r)\mathrm{d}r=\frac{1}{s^{\alpha_1+1}(1-s)^{\alpha_1+1}} \frac{r^{2\alpha_1+1}}{2^\alpha \Gamma(\alpha+1)} e^{-\frac{r^2}{2}\left( \frac{1}{s}+\frac{1}{1-s}\right)} \mathrm{d}r.
\end{align}
Integrating this expression against $r^{2\alpha_1-2}$ yields the previously claimed expectation value
\begin{align}
    C_{\alpha_1} := \mathbb{E}\left[ \int_0^1 |X(t)|^{2\alpha_1-2} \right] = \int_0^1 \mathrm{d}s \int_0^\infty \mathrm{d}r\,r^{2\alpha_1-2} \rho_{s}(r) = 2^{\alpha_1-1} \frac{\Gamma(\alpha_1)}{\alpha_1}.
\end{align}

Suppose now that $\hat{X}(t)$ is the biased Brownian excursion and, conditionally on $\hat{X}(t)$, let $U \in [0,1]$ be a random variable sampled with density proportional to $|\hat{X}(u)|^{2\alpha_1-2}\mathrm{d}u$.
Then we let
\begin{equation}\label{eq:Trelations}
    T_1 = \frac{U}{|\hat{X}(U)|^2},\qquad T_2 = \frac{1-U}{|\hat{X}(U)|^2}, \qquad S = \frac{\hat{X}(U)}{|\hat{X}(U)|}.
\end{equation}
We will demonstrate that $T_1$, $T_2$ and $S$ are independent and $T_1$ and $T_2$ are distributed precisely as the inverse gamma distribution mentioned above. 

From \eqref{eq:excdistancedist} it follows that the joint distribution of the pair $(|\hat{X}(U)|, U)$ has probability density
\begin{align}
    \frac{1}{C_{\alpha_1}} \rho_{u}(r) r^{2\alpha_1-2} \mathrm{d}r\mathrm{d}u \qquad \text{on } \mathbb{R}_{>0}\times (0,1).
\end{align}
Since $T_1$ and $T_2$ are bijectively related to $|\hat{X}(U)|$ and $U$ via \eqref{eq:Trelations}, the joint density of the pair $(T_1,T_2) \in \mathbb{R}_{>0}^2$ is obtained from this by the transformation $r = 1/\sqrt{\tau_1+\tau_2}$, $u = \tau_1 / (\tau_1+\tau_2)$,  with Jacobian $2 r^{-5}\mathrm{d}r\mathrm{d}u = \mathrm{d} \tau_1 \mathrm{d} \tau_2$, which yields
\begin{align}
    \frac{1}{C_{\alpha_1}} \rho_{u}(r) r^{2\alpha_1-2} \mathrm{d}s\mathrm{d}r = \frac{1}{(2^{\alpha_1} \Gamma(\alpha_1))^2}\tau_1^{-\alpha_1-1} e^{-\frac{1}{2\tau_1}} \tau_2^{-\alpha_1-1} e^{-\frac{1}{2\tau_2}} \mathrm{d}\tau_1 \mathrm{d}\tau_2.
\end{align}
Comparing with \eqref{eq:invgamma}, we observe that $T_1$ and $T_2$ are independent and distributed with the desired inverse gamma distribution.

Finally, conditionally on $T_1$, $T_2$ and $S$, the curves $\hat{X}(U-t)$ and $\hat{X}(U+t)$ are independent $d$-dimensional Brownian motions both started at $\hat{X}(U) = S / \sqrt{T_1+T_2}$ and conditioned to hit to the origin after time $U = T_1 / (T_1+T_2)$ and $1-U = T_2 / (T_1+T_2)$ respectively.
By the scale invariance of the Brownian motion the curves 
\begin{equation}
    X^{1}(t) = \sqrt{T_1+T_2} \hat{X}\left(\frac{T_1 - t}{T_1+T_2}\right), \qquad X^{2}(t) = \sqrt{T_1+T_2} \hat{X}\left(\frac{T_1 + t}{T_1+T_2}\right)
\end{equation}
are distributed as independent $d$-dimensional Brownian motions started at $S$ and conditioned to hit the origin after time $T_1$ and $T_2$ respectively.
But as we computed above, $T_1$ and $T_2$ have precisely the distribution of the hitting time of the origin of a $d$-dimensional Brownian motion started at $S$ and conditioned to tough the boundary $\partial W$ at the origin, so we may lift the latter conditioning.
Since this is precisely the inverse of \eqref{eq:bmconcat}, we have proven that the identity \eqref{eq:bmconcat} holds for the law of $\hat{X}(t)$.

\subsection{Brownian motion conditioned to hit the origin}

We have seen that the biased excursion $\hat{X}(t)$, and therefore also the biased Mated-CRT graph $\hat{G}_n^{\mathbf{C}}$, can be constructed from a pair of $d$-dimensional Brownian motions that are conditioned to hit the boundary $\partial W$ at the origin.
Let us discuss these processes in a bit more detail.
An important role is played by the harmonic function 
\begin{equation}
    h(x) = |x|^{-\alpha_1-\frac{d}{2}+1} m_1\left(\frac{x}{|x|}\right).
\end{equation}
It can be recovered from the heat kernel by first computing the $t$-integral
\begin{equation}
    \int_0^\infty P_t^{\mathbf{C}}(x,y) \mathrm{d}t = \sum_{j=1}^\infty \frac{1}{\alpha_j} |x|^{1-\frac{d}{2}-\alpha_j} |y|^{1-\frac{d}{2}+\alpha_j}  m_j\left(\frac{x}{|x|}\right)m_j\left(\frac{y}{|y|}\right)\qquad \text{for }|x|>|y|.
\end{equation}
As $y$ tends to zero we thus have
\begin{equation}
    \int_0^\infty P_t^{\mathbf{C}}(x,y) \mathrm{d}t \stackrel{y\to 0}{\sim}  h(x)\, h(y) \frac{|y|^{2\alpha_1}}{\alpha_1},
\end{equation}
which estimates the probability that an unrestricted $d$-dimensional Brownian motion started at $x$ leaves the cone $W$ at a point close to the origin.
Even though the Brownian motion has vanishing probability of hitting $\partial W$ at the origin, we may still condition on this event by a so-called Doob's $h$-transform \cite{doob1984classical} of a standard Brownian motion with respect to this harmonic function.
Without diving into the theory of $h$-transforms, we can characterize the Brownian motion $X^i(t)$ conditioned to hit $\partial W$ at the origin as follows.
If $A \subset W$ is a closed neighbourhood of $x_0 = X^i(0)$ and we consider the exit time $\tau$ when $X^i(t)$ leaves $A$, then the distribution of the exit point $X^i(\tau)$ is related to that of a standard Brownian motion started at $x_0$ by a factor $h(X^i(\tau))/h(x_0)$.

\begin{figure}[H]
  \centering
    \includegraphics[width=0.45\textwidth]{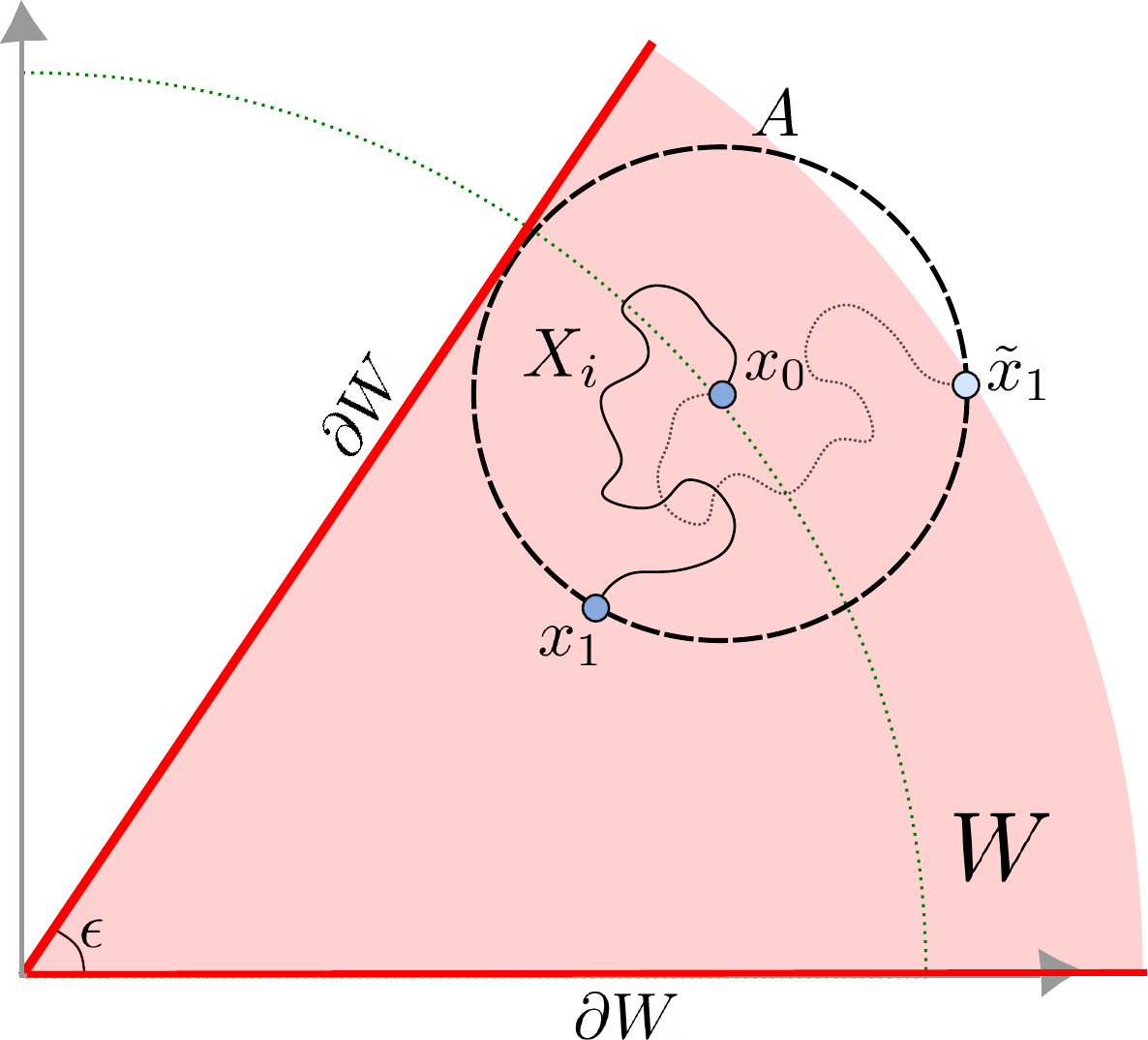}
  \caption{The first step in the iterative procedure to produce $X^{i}(t)$ in the case of a two-dimensional cone $W$: $x_0$ is sampled from the unit circle $W \cap \mathbb{S}^{1}$ with density $m_1(x)^2 = h(x)^2$; a standard 2-dimensional Brownian motion is run until it exits the disk $A$ at $\tilde{x}_1$, which is then rotated to end at the random position $x_1$ with distribution $h(x)/h(x_0)$. This procedure is to be repeated with a new disk centered at $x_1$.}
  \label{fig:sampling}
\end{figure}

This characterization gives a simple iterative procedure of constructing $X^i(t)$ started at $x_0 \in W$ from standard Brownian motion, see Figure~\ref{fig:sampling}. 
We take the subset $A$ to be the largest Euclidean ball centered at $x_0$ and contained in $W$ and let $x_1$ be a random variable on the sphere $\partial A$ with probability density $h(x) / h(x_0)$.
We may then consider a standard Brownian motion started at $x_0$ until it hits the boundary of the ball at time $t_1$. 
By symmetry this happens at a uniform point $\tilde{x}_1$ on $\partial A$.
A $d$-dimensional rotation around $x_0$ that brings $\tilde{x}_1$ to $x_1$ then gives an appropriately sampled path for $X^i(t)$ for $t\in [0,t_1]$.
Since $X^{i}(t)$ is a Markov process, we may iterate this procedure with the new starting point $X^{i}(t_1) = x_1$ to obtain the path for $X^{i}(t)$ with $t\in [t_1,t_2]$, etcetera.
Of course, infinite iteration is required to reach the origin, but if one is only interested in the path until reaching some small distance $\epsilon>0$ from the origin, then the number of required iterations can be seen to grow only logarithmically in $1/\epsilon$.

\section{Simulations and Hausdorff dimension estimates}\label{section_numerical}

\subsection{Sampling Mated-CRT graphs numerically}

In the previous section we have introduced the random Mated-CRT graph $\hat{G}_n^{\mathbb{C}}$ constructed from the biased Brownian excursion.
Let us now turn to the numerical implementation of this construction.

Sampling $\hat{G}_n^{\mathbb{C}}$ with exactly the right probability distribution is challenging, as it relies on continuous Brownian motions.
What helps is that the graph $\hat{G}_n^{\mathbb{C}}$ is determined by the ranges of $X^1(t)$ and $X^2(t)$ on intervals of length $(T_1+T_2)/n$, where $T_1$ and $T_2$ are the time extents that we know to be inverse-gamma distributed. 
The probability of $T_1$ or $T_2$ being much shorter than their expectation value $1/(2\alpha-2)$ is very small.
Hence, to approximate $\hat{G}_n^{\mathbb{C}}$ well, it suffices to sample the Brownian motions at a time resolution $\epsilon$ that is significantly smaller than $\mathbb{E}[(T_1+T_2)/n]= 1/(n(\alpha-1))$.
This we do by approximating the Brownian motion by a driftless random walk with increments that are sampled uniformly on the sphere of radius $\sqrt{\epsilon}$ in $\mathbb{R}^d$. In this case, we use $\sqrt{\epsilon}\in[0.0001,0.001]$.
The reason to opt for these increments instead of the potentially more accurate Gaussian increments is that the exit times and exit positions (e.g. from the cone $W$) are more easily controlled with bounded increments.

To be precise, for a desired correlation matrix $\mathbf{C}$ we compute the exponent $\alpha_1$ and fundamental mode $m_1$ (either analytically or numerically if an analytical solution is not available). 
Then to obtain a single sample of $\hat{G}_n^{\mathbb{C}}$ we perform the following procedure, based on the construction in Section~\ref{sec:biasedbm}.
A random starting point $x_0$ with distribution $m_1(x)^2$ on $W \cap \mathbb{S}^{d-1}$ is chosen using rejection sampling.
Two random piece-wise linear curves from $x_0$ to the origin are obtained by running the random walk in an iterative fashion as follows.
We find the largest radius $r$ such that the ball $\mathrm{Ball}_{r}(x_0)$ around $x_0$ is contained in $W$ and choose a point $x_1$ on its boundary with distribution $h(x)/h(x_0)$, again using rejection sampling.
Next, we run the mentioned random walk with steps of size $\sqrt{\epsilon}$ until it leaves $\mathrm{Ball}_{r}(x_0)$, denoting the exit point on the sphere by $\tilde{x}_1$.
This random walk is rotated by an orthogonal transformation that only depends on $x_0$, $x_1$ and $\tilde{x}_1$ to produce a piece-wise linear path from $x_0$ to $x_1$ (note that the last segment of this path has to be shortened a bit to end precisely at $x_1$ instead of ending outside $\mathrm{Ball}_{r}(x_0)$).
We iterate this procedure, but now using $x_1$ as the starting point, which extends the piece-wise linear path from $x_1$ to a random point $x_2$ on the boundary of the largest ball $\mathrm{Ball}_{r}(x_1)$ around $x_1$, and so on.
This is continued until we reach a point within distance $2\sqrt{\epsilon}$ from the origin, after which we add a final segment connecting to the origin.
The result is a piece-wise linear path from $x_0$ to the origin that stays strictly in the cone $W$ and approximates the law of the Brownian motion in $W$ conditioned to hit the origin.
Concatenating the two paths according to \eqref{eq:bmconcat} leads to a piece-wise linear excursion that approximates the biased unit-time Brownian excursion $\hat{X}(t)$.
Finally, the mated-CRT graph of size $n$ is obtained as explained in Section \ref{sec:matedcrtgraphs}, resulting in an adjacency matrix for the $n$ vertices of the graph.

\subsection{Hausdorff dimension estimates via finite-size scaling}\label{section:results}\label{subsection:Finite-size}

As explained in Section \ref{sec:matedcrtgraphs}, the central question is whether the metric space induced by the graph distance on $\hat{G}_n^{\mathbf{C}}$ possesses a scaling limit. 
Does there exist a real number $d_H^{\mathbf{C}}>0$, which we then call the Hausdorff dimension of the model, such that the metric space $n^{-1/d_H^{\mathbf{C}}}\hat{G}_n^{\mathbf{C}}$ has a limit as $n\to\infty$ (in the Gromov--Hausdorff sense)?
This statement about the limit is not something one can effectively verify numerically, but there is a necessary condition that is within numerical reach.
Let $d_n$ be the graph distance between two uniformly sampled vertices in the random graph $\hat{G}_n^{\mathbf{C}}$.
Then for the existence of a sane Gromov--Hausdorff limit it is necessary that $d_n/n^{1/d_H^{\mathbf{C}}}$ converges in distribution as $n\to\infty$.
Since $d_n$ is relatively easy to measure, this allows us to verify the convergence in distribution and at the same time estimate the value $d_H^{\mathbf{C}}$ through finite-size scaling.

\begin{figure}[H]
  \centering
    \includegraphics[width=0.8\textwidth]{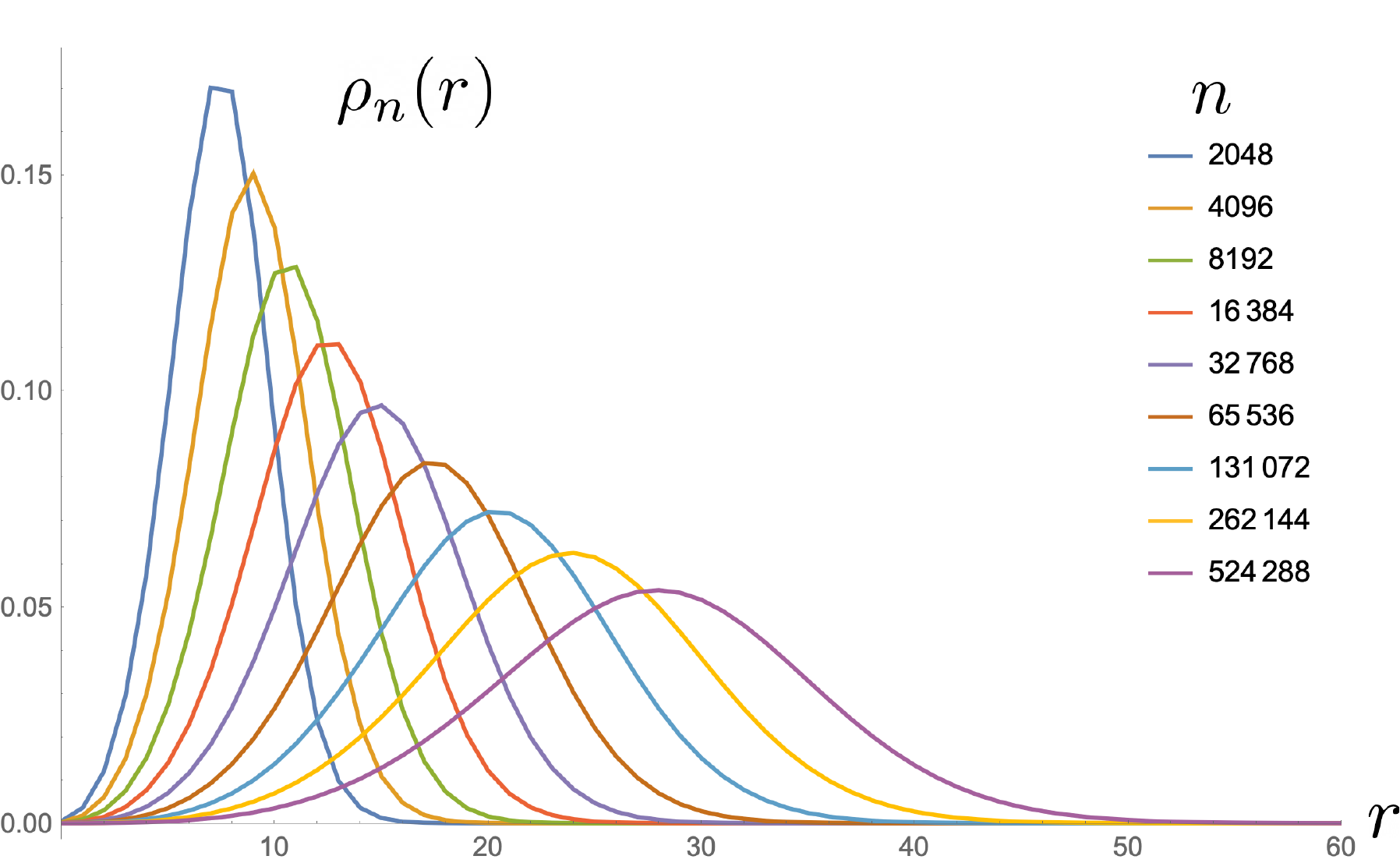}
  \caption{Normalized histogram $\rho_n(r)$ for $\alpha=\beta=\gamma=\frac{\pi}{2}$, obtained by sampling $m=10000$ graphs independently and measured distances $k=10$ times from a uniformly random chosen vertex.}
  \label{fig:hist_original}
\end{figure}

The probability distributions $\rho_n(r) = \mathbb{P}(d_n=r)$ for $r=0,1,2,\ldots$ were estimated for $n = 2^{11}, 2^{12}, \ldots, 2^{19}$ as follows. 
For each size $n$, the graph $\hat{G}_n^{\mathbf{C}}$ was sampled $80000$ times and, in each sampled graph, the graph distances from a uniformly chosen vertex to all other vertices were determined several times.
All these distances were stored in a histogram, which upon normalization, provides our best estimate for $\rho_n(r)$ with small statistical errors.
See Figure \ref{fig:hist_original}.

For convenience we extend $\rho_n(r)$ to a continuous function of $r\in\mathbb{R}_{\geq 0}$ via linear interpolation.
If $d_n / n^{1/d_H^{\mathbf{C}}}$ converges in distribution to a random variable with density $\rho(x)$, we expect to have the limit  
\begin{equation}
    \underset{n\rightarrow\infty}{\mathrm{lim}}n^{1/d_H}\rho_n(n^{1/d_H}x)=\rho(x).\label{limitrho}
\end{equation}
This a mildly stronger assumption than what is implied by the Gromov--Hausdorff convergence, but one that is supported by our data.
In order to study the limit \eqref{limitrho}, we choose as reference size $n_0=2^{19}$ and aim to collapse the curves $\rho_n$ for the other sizes $n$ to $\rho_{n_0}$. 
More precisely, for each $n=2^{11},\ldots,2^{18}$ we determine fit parameters $k_n$ and $s_n$ that minimize the integrated square deviation between $k_n^{-1}\rho_n(k_n^{-1}(x+s_n)-s_n)$ and $\rho_{n_0}$. 
The shift $s_n$ is included to compensate for discretization effects and is largely independent of $n$.
By comparing this expression with \eqref{limitrho}, we see that $k_n\sim C n^{-1/d_H}$, i.e., finding the asymptotic behaviour of $k_n$ is the key to estimating $d_H$. 

\begin{figure}[H]
  \centering
    \includegraphics[width=0.8\textwidth]{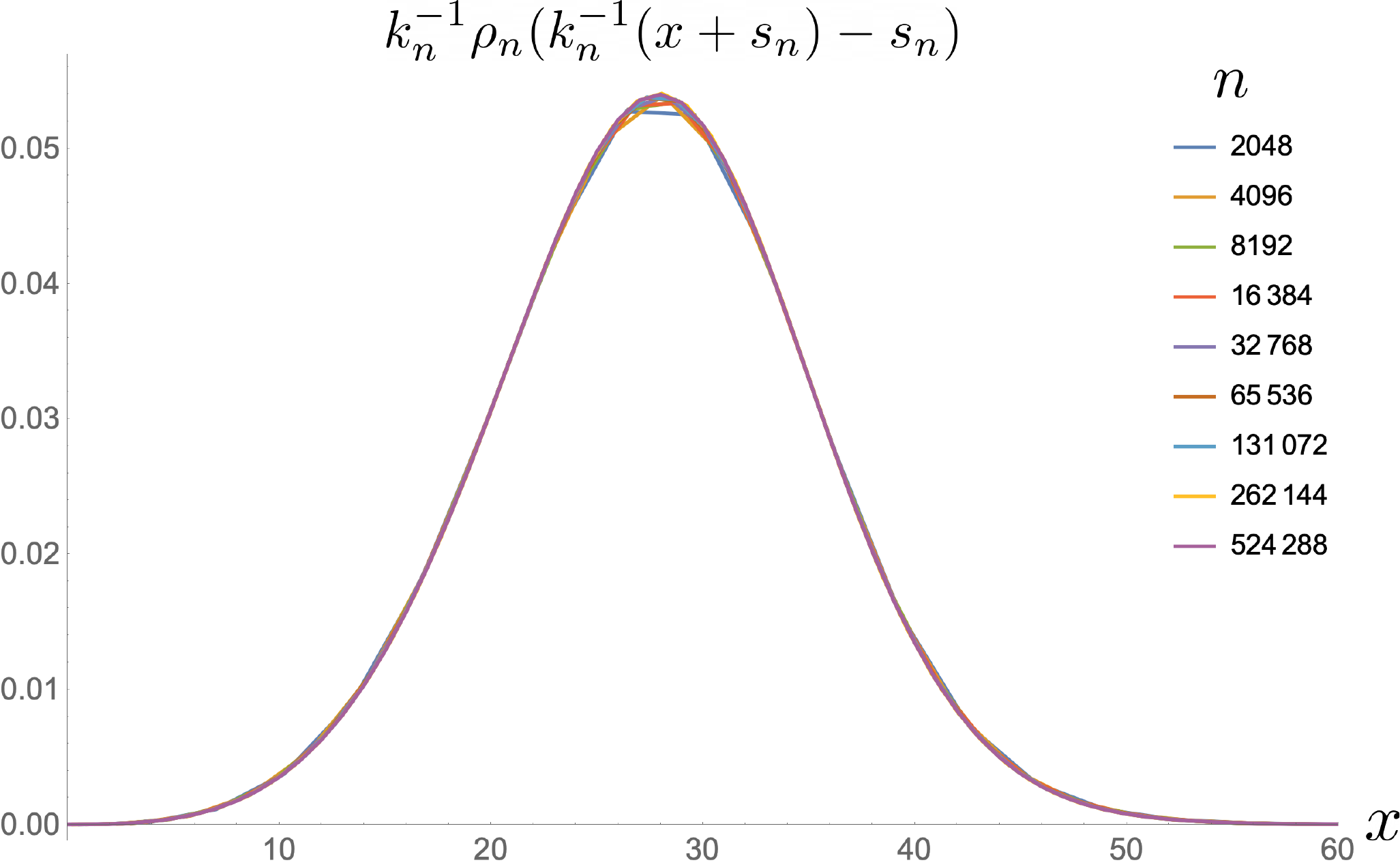}
  \caption{Collapsed histograms with optimal shift for $\alpha=\beta=\frac{\pi}{2}$ and $n_0=2^{19}$.}
  \label{fig:hist_collapse}
\end{figure}
In order to find more accurate values for $d_H$, we collapse $\rho_n$ two times. In the first one $k_n^{-1}\rho_n(k_n^{-1}(x+s_n)-s_n)\longmapsto\rho_{n_0}$, we extract the values ${s_n}$ to compute its mean $s$. In the second one, we use $s$ to collapse $k_n^{-1}\rho_n(k_n^{-1}(x+s)-s)\longmapsto\rho_{n_0}$ and we extract $k_n$. Finally, we estimate $d_H$ by fitting $k_n$ to the ansatz
\begin{equation}
   \left(\frac{n}{n_0}\right)^{-1/d_H}\left(a+b\left(\frac{n}{n_0}\right)^{-\delta}\right),
\end{equation}
where $a\approx 1$, $\delta$ of order $1/d_H$ and $|b|\ll 1$.
This expression takes into account a leading-order correction and has proven to work well for Hausdorff dimension estimations in a similar setting \cite{Barkley_2019}. The fitting procedure was tested by varying the range of volumes included, as well as the values of $\delta$ and $d_H$ while keeping $a\approx 1$ and $|b|\ll 1$. In this way, we determine systematic errors.
On the other hand, the statistical errors in the fit parameters were determined using batching (by dividing the data into eight independent batches and applying the analysis each independently).
The results are presented in the next subsections.

\subsection{Results for Mated-CRT maps ($d=2$)}

\begin{figure}[H]
  \centering
    \includegraphics[width=0.7\textwidth]{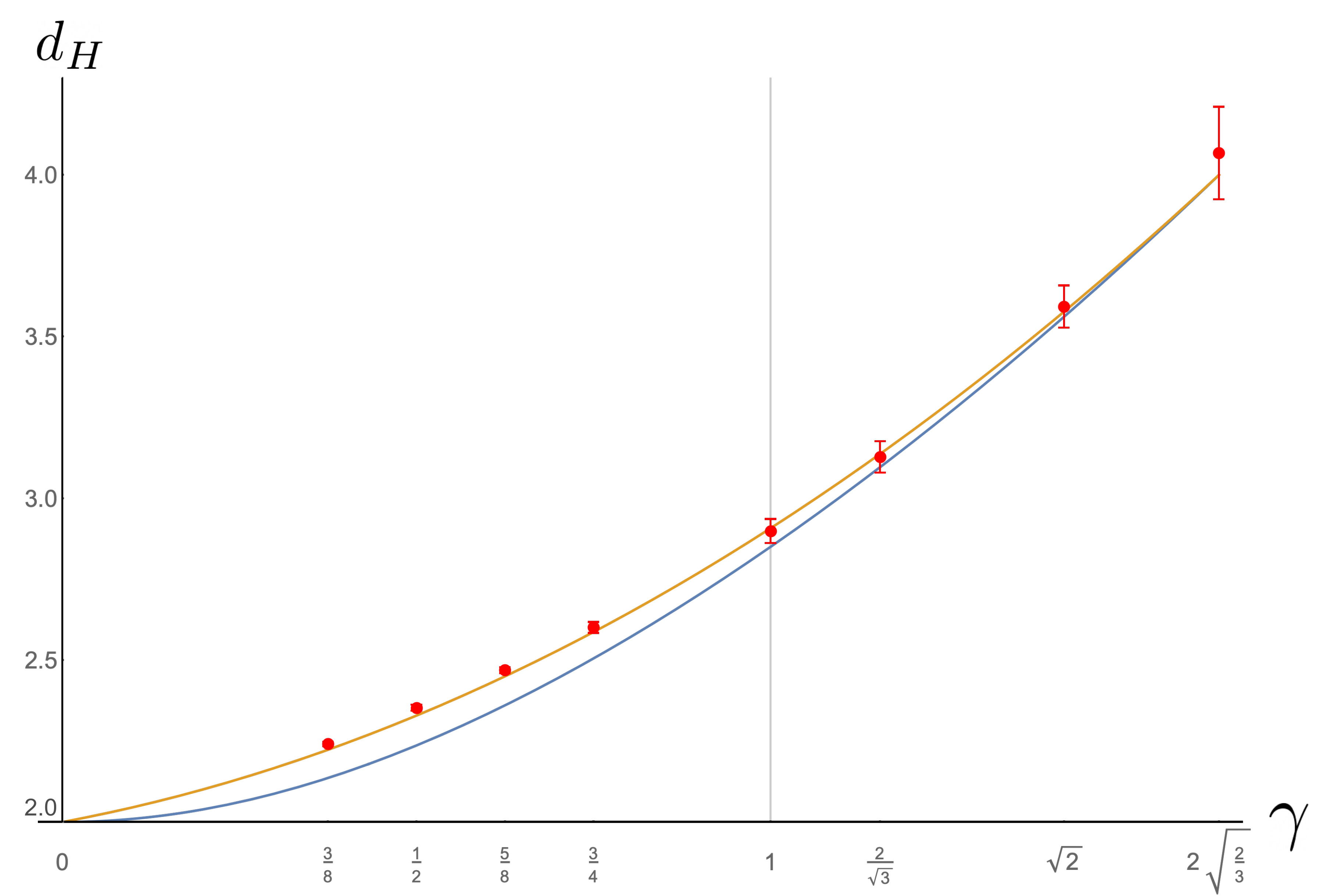}
  \caption{Hausdorff dimension estimates from simulated mated-CRT maps for different values of $\gamma$. Watabiki's formula \eqref{dH_watabiki} is plotted in blue, while Ding and Gwynne's formula \eqref{dH_DG} is plotted in yellow.}
  \label{fig:dH_d2}
\end{figure}

As explained in Section~\ref{subsection:matedcrt}, the Gromov--Hausdorff convergence of the Mated-CRT maps has not been proved, but the number of vertices within a ball of radius $r$ in the graph $\hat{G}_n^{\mathbf{C}}$ with very large $n$ is known to grow as $r^{d_{\gamma}}$ as $r\to\infty$, where $d_{\gamma}$ is the Hausdorff dimension of Liouville Quantum Gravity.
Here the covariance $\mathbf{C}_{12}=-\cos\left(\alpha\right)$ is related to $\gamma$ through the relation $\alpha=\pi\gamma^2/4$.
This strongly suggests that the convergence \eqref{limitrho} holds with Hausdorff dimension $d_H^{\mathbf{C}} = d_{\gamma}$, and thus our methods provide a means of estimating the Hausdorff dimension $d_\gamma$ of Liouville Quantum Gravity.
In Figure \ref{fig:dH_d2} and Table \ref{table:dH_2D}, we show the numerical values of $d_H$ as a function of $\gamma$. 

A distinction between the regions $\gamma<1$ and $\gamma\geq1$ is made, since the latter is the domain analysed in \cite{Barkley_2019}. The reason for choosing the four values $\gamma=1,\sqrt{4/3},\sqrt{2},\sqrt{8/3}$ is that they are the values associated to the universality classes of Schnyder-wood-decorated triangulations, bipolar-oriented triangulations, spanning-tree-decorated quadrangulations and uniform quadrangulations, respectively.
For these models, discrete mating-of-trees bijections are available that are at the basis of the high-precision estimates of $d_H$ in \cite{Barkley_2019}.
They thus form a good benchmark for the techniques developed in this work.
Our results in Table~\ref{table:dH_2D} are seen to be very well consistent with those in \cite[Table 5]{Barkley_2019}, although the errors here are significantly larger.

Our current method has the advantage that it can be used to perform simulations at  any $\gamma\in(0,2)$, in particular in the region $\gamma<1$ where very few numerical estimates for $d_H$ were known (see \cite[Section 5]{Barkley_2019} for estimates based on Liouville first-passage percolation). 
Gaining more accurate estimates for small $\gamma$ is important, because it is in this region that some proposed formulas for the Hausdorff dimension deviate substantially. 
Two such formulae are the one due to Watabiki \cite{Watabiki:1993fk} (shown in blue in Figure \ref{fig:dH_d2})
\begin{equation}
    d^{W}=1+\frac{\gamma^2}{4}+\sqrt{\left(1+\frac{\gamma^2}{4}\right)^2+\gamma^2},
    \label{dH_watabiki}
\end{equation}
and one by Ding and Gwynne \cite{Ding:2018uez} (shown in yellow in Figure \ref{fig:dH_d2})
\begin{equation}
   d^{DG}=2+\frac{\gamma^2}{2}+\frac{\gamma }{\sqrt{6}}.
   \label{dH_DG}
\end{equation}
As can be seen in Figure \ref{fig:dH_d2}, our new estimates strengthen the conclusion of \cite{Barkley_2019} that Watabiki's formula is ruled out numerically (in addition to being already inconsistent with the $\gamma\to 0$ bounds in \cite{ding2019upper}).
However, the measurements are still statistically consistent with Ding and Gwynne's formula.
\begin{table}[H]
    \centering
    \begin{tabular}{|c|c|}
    \cline{1-2}
            $\gamma$ & $d_H$ \\
           \hline
           $3/8$  &  $2.24\pm 0.01$ \\
          \hline
           $1/2$  &  $2.35\pm 0.01$ \\
          \hline
           $5/8$  &  $2.47\pm 0.01$ \\
          \hline
          $3/4$  &  $2.60\pm 0.02$ \\
          \hline
          $1$  &  $2.90\pm 0.04$ \\
          \hline
          $\sqrt{4/3}$ & $3.13\pm 0.05$ \\
          \hline
          $\sqrt{2}$ & $3.59\pm 0.07$ \\
          \hline
          $\sqrt{8/3}$ & $4.07\pm 0.14$ \\
          \hline
    \end{tabular}
    \caption{Our Hausdorff dimension estimates from simulated mated-CRT maps for different values of $\gamma$. The errors have been determined according to the procedure outlined at the end of Section~\ref{subsection:Finite-size}.}
    \label{table:dH_2D}
\end{table}

\subsection{Results for Mated-CRT graphs in $d=3$}
\begin{figure}[H]
  \centering
    \includegraphics[width=0.3\textwidth]{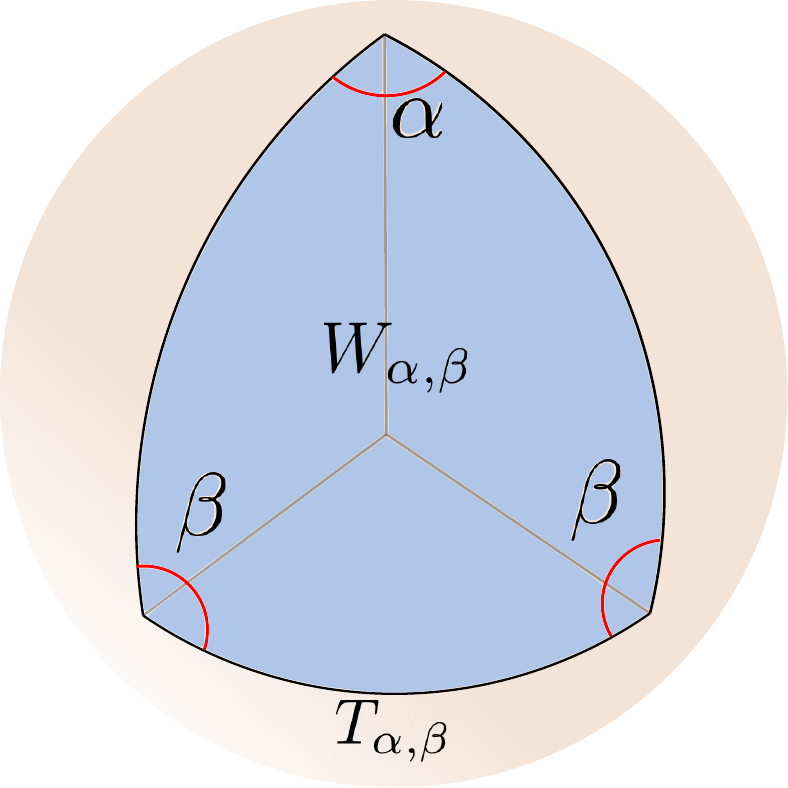}
  \caption{The cone $W_{\alpha,\beta}$ is spanned by the isosceles spherical triangle $T_{\alpha,\beta}$ in the unit sphere.}
  \label{fig:3D_wedge_iso}
\end{figure}
Having benchmarked the numerical methods, we turn to the main numerical results of this work, the Hausdorff dimension estimates of mated-CRT graphs in $d=3$. 
Since the phase diagram is significantly larger than in $d=2$, being three-dimensional instead of one-dimensional, we have chosen to restrict our attention to the two-dimensional subspace corresponding to isosceles spherical triangles (Figure~\ref{fig:3D_wedge_iso}) in which two of the angles are equal: $\gamma = \beta$ or, equivalently, $\mathbf{C}_{23}=\mathbf{C}_{13}$\footnote{These results extend of course to the planes $\alpha=\gamma$ and $\alpha=\beta$ due to rotational symmetry.}.

\begin{figure}[]
  \centering
  \includegraphics[width=1\textwidth]{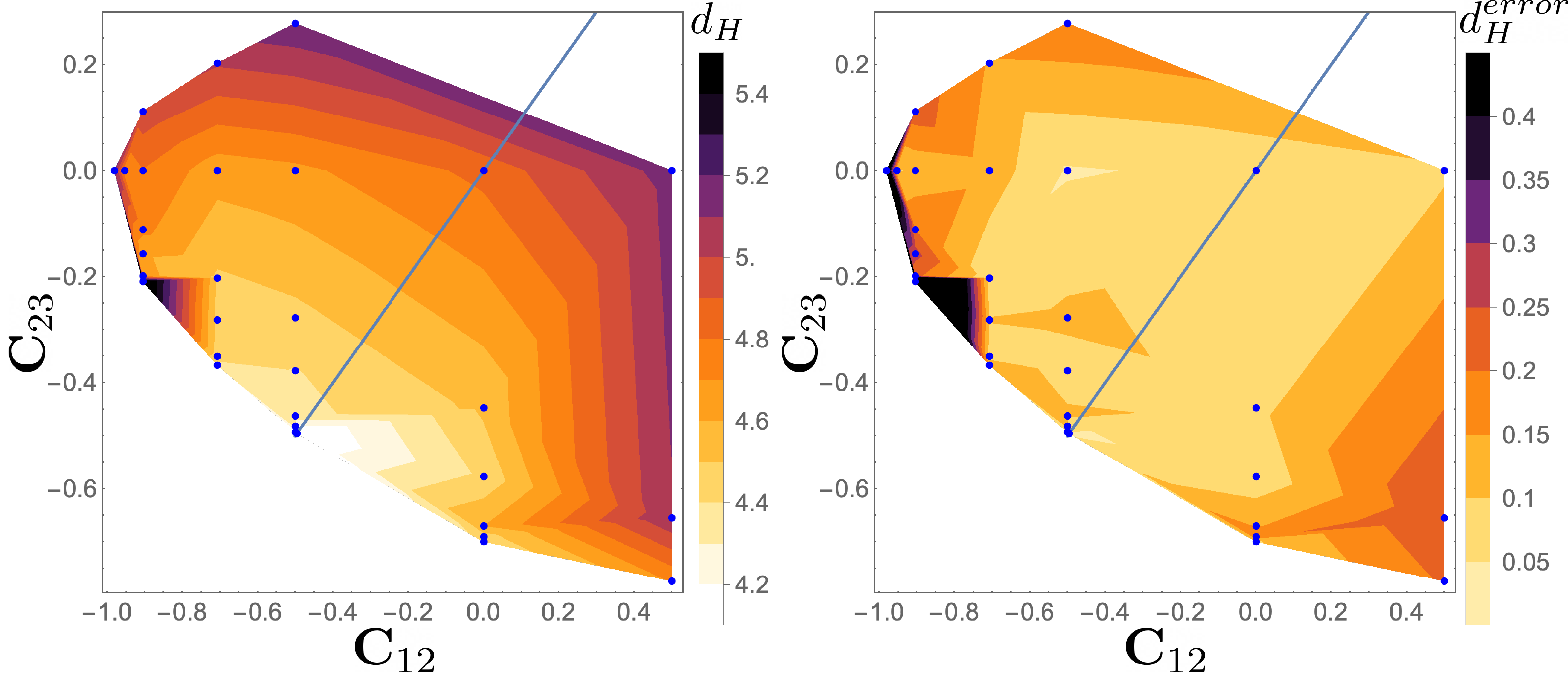}
    
  \includegraphics[width=0.6\textwidth]{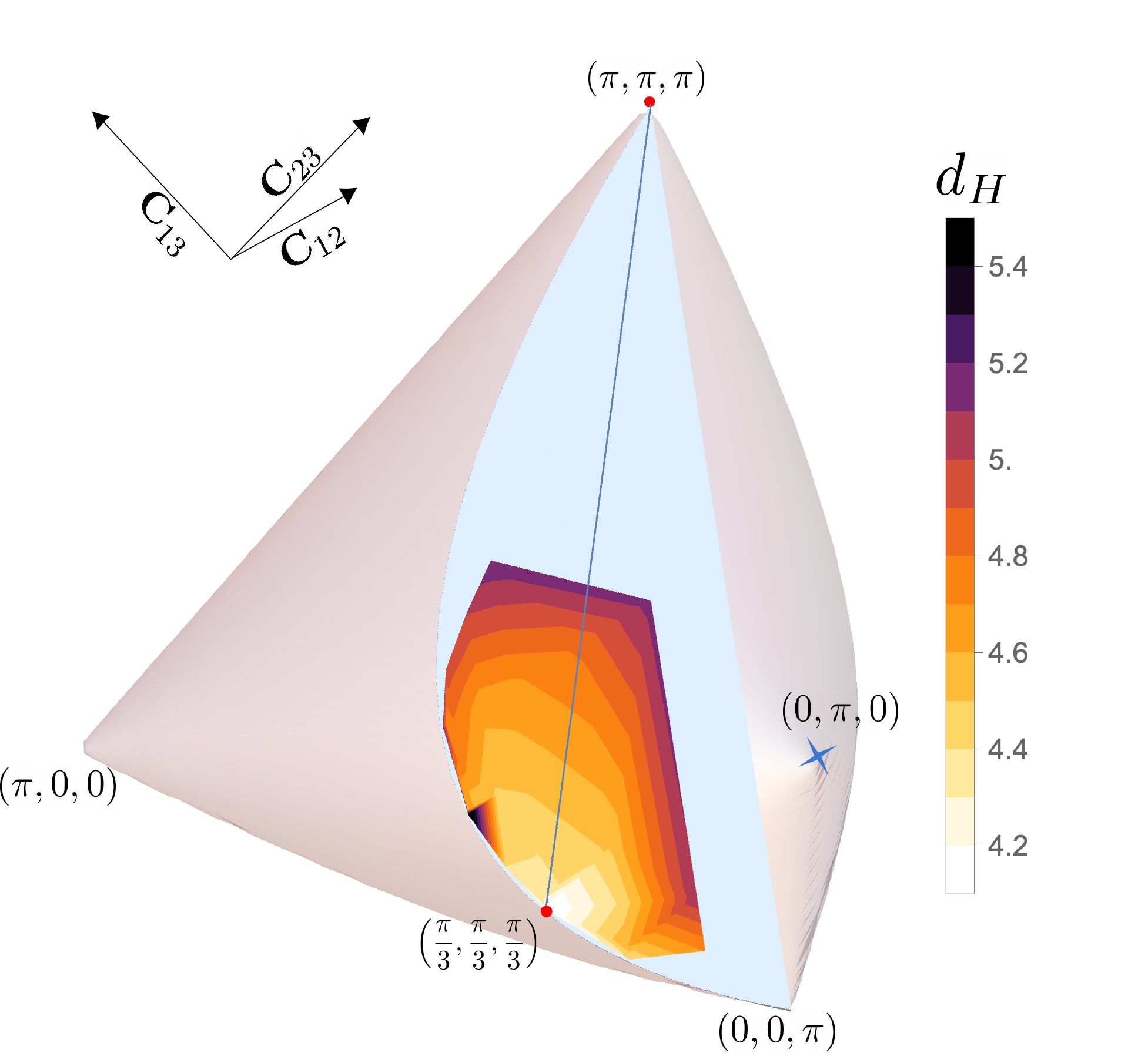}
  \caption{Top left: Hausdorff dimension estimates from Mated-CRT graphs constructed from correlated 3D Brownian Excursions as a function of the covariances $\mathbf{C}_{12}=-\cos{(\alpha)}$ and $\mathbf{C}_{23}=\mathbf{C}_{13}=-\cos{(\beta)}$. The contours are based on a linear interpolation of the simulated data points (see Table~\ref{table:dH_3D}) that are shown in blue. The light blue line ($\mathbf{C}_{12}=\mathbf{C}_{23}$) indicates the models corresponding to equilateral spherical triangles. Top right: The corresponding errors in the estimates (including both systematic and statistical contributions). Bottom: The same contours as the top left plot shown in the full three-dimensional phase diagram.}
  \label{fig:dH_d3}
\end{figure}

The estimates for the Hausdorff dimension $d_H^{\mathbf{C}}$ including error bars for a variety of angle pairs $(\alpha,\beta)$ are presented in Figure~\ref{fig:dH_d3} and listed in Table \ref{table:dH_3D}. For convenience we record a reasonable fit for $d_H^{\mathbf{C}}$ using a quadratic ansatz in $\mathbf{C}$ that respects the symmetries,
\begin{equation}\label{extrapolation}
d_H \approx 4.83 +0.42 (\mathbf{C}_{12}+\mathbf{C}_{23}+\mathbf{C}_{13}) + 0.37 (\mathbf{C}_{12}^2+\mathbf{C}_{23}^2+\mathbf{C}_{13}^2) -0.38 (\mathbf{C}_{12}\mathbf{C}_{23}+\mathbf{C}_{13}\mathbf{C}_{23}+\mathbf{C}_{12}\mathbf{C}_{23}).
\end{equation}

Although we have only been able to effectively simulate a limited region of the full phase diagram, several conclusions can be drawn based on the data. 
First of all, the dependence of $d_H^{\mathbf{C}}$ on $\mathbf{C}$ differs qualitatively from that of the string susceptibility in Figure~\ref{fig:cov_tetrahedron}, suggesting that we are really dealing with a multi-parameter family of universality classes. 
Secondly, contrary to the two-dimensional case there appears to be no limit, at least in the isosceles region, where the Hausdorff dimension approaches a ``classical'' value equal to $d$ itself, in this case $3$. 
Instead we seem to observe a minimum $d_H \approx 4.1$ when $(\alpha,\beta,\gamma)\rightarrow(\pi/3,\pi/3,\pi/3)$ corresponding to a Brownian excursion in the cone spanned by a tiny equilateral spherical triangle. 
\begin{table}[H]
    \centering
    \begin{tabular}{cccc}
    \hline
             $(\alpha,\beta)$ & $d_H$ & $(\alpha,\beta)$ & $d_H$\\ [0.5ex]
           \hline
           $\left(\frac{\pi }{16},\frac{\pi }{2}\right)$  &  $5.12\pm 0.55$             & $\left(\frac{\pi }{3},\frac{\pi }{3}+0.04\right)$  &  $4.36\pm 0.11$\\ [0.3ex]
           $\left(\frac{25\pi }{256},\frac{\pi }{2}\right)$  &  $4.72\pm 0.09$ &$\left(\frac{\pi }{3},\frac{\pi }{3}+0.80\right)$  &  $5.18\pm 0.19$ \\
           [0.3ex]
           $\left(\frac{9\pi }{64},1.14\right)$  &  $5.61\pm 1.16$ &            $\left(\frac{\pi }{3},1.18\right)$  &  $4.39\pm 0.08$ \\
           [0.3ex]
           $\left(\frac{9\pi }{64},1.37\right)$  &  $4.63\pm 0.21$ & $\left(\frac{\pi }{3},1.29\right)$  &  $4.47\pm 0.11$ \\
           [0.3ex]
           $\left(\frac{9\pi }{64},1.41\right)$  &  $4.61\pm 0.26$ &  $\left(\frac{\pi }{3},\frac{\pi }{2}\right)$  &  $4.67\pm 0.04$\\
           [0.3ex]
           $\left(\frac{9\pi }{64},1.46\right)$  &  $4.73\pm 0.20$ & $\left(\frac{\pi }{3}+0.005,\frac{\pi }{3}+0.005\right)$  &  $4.08\pm 0.04$\\
           [0.3ex]
           $\left(\frac{9\pi }{64},\frac{\pi }{2}\right)$  &  $4.79\pm 0.11$ &  $\left(\frac{\pi }{2},0.79\right)$  &  $4.42\pm 0.09$ \\
           [0.3ex]
           $\left(\frac{9\pi }{64},1.68\right)$  &  $4.96\pm 0.23$ & $\left(\frac{\pi }{2},0.81\right)$  &  $4.45\pm 0.18$\\
           [0.3ex]
           $\left(\frac{\pi }{4},1.19\right)$  &  $4.36\pm 0.13$ &$\left(\frac{\pi }{2},0.83\right)$  &  $4.60\pm 0.14$  \\
           [0.3ex]
           $\left(\frac{\pi }{4},1.21\right)$  &  $4.45\pm 0.08$ & $\left(\frac{\pi }{2},0.95\right)$  &  $4.42\pm 0.06$ \\
           [0.3ex]
           $\left(\frac{\pi }{4},1.28\right)$  &  $4.45\pm 0.11$ & $\left(\frac{\pi }{2},1.11\right)$  &  $4.52\pm 0.09$ \\
           [0.3ex]
           $\left(\frac{\pi }{4},1.37\right)$  &  $4.49\pm 0.06$ & $\left(\frac{\pi }{2},\frac{\pi }{2}\right)$  &  $4.83\pm 0.06$\\
           [0.3ex]
           $\left(\frac{\pi }{4},\frac{\pi }{2}\right)$  &  $4.64\pm 0.13$ & $\left(\frac{2\pi }{3},0.68\right)$  &  $4.71\pm 0.21$ \\
           [0.3ex]
           $\left(\frac{\pi }{4},1.77\right)$  &  $5.01\pm 0.14$ &$\left(\frac{2\pi }{3},0.86\right)$  &  $5.09\pm 0.25$ \\
          [0.3ex]
          $\left(\frac{\pi }{3},\frac{\pi}{3}+0.01\right)$  &  $4.12\pm 0.04$ & $\left(\frac{2\pi }{3},\frac{\pi }{2}\right)$  &  $5.15\pm 0.09$ \\
          [0.3ex]
          $\left(\frac{\pi }{3},\frac{\pi }{3}+0.02\right)$  &  $4.35\pm 0.11$ &    &\\
          [0.3ex]
          \hline
    \end{tabular}
    \caption{Hausdorff dimension measurements with error bars of the Mated-CRT graph $\hat{G}_n^{\mathbf{C}}$ obtained from correlated 3D Brownian Excursions. The angles $(\alpha,\beta)$ correspond to the spherical angles of the isosceles spherical triangles $T_{\alpha,\beta}$.}
    \label{table:dH_3D}
\end{table}
\section{Discussion}\label{section_discussion}

In this study, we have proposed a sequence of discrete metric spaces $G_n^{\mathbf{C}}$, Mated-CRT graphs, associated to a correlated Brownian excursion in $d$ dimensions, generalizing the Mated-CRT maps in $d=2$.
We hypothesize that upon normalization of distances these metric spaces approach a non-trivial continuous random metric as $n\to\infty$ that inherits its scaling properties from the Brownian excursion.
In $d=2$ this has largely been demonstrated as part of the mating of trees approach to Liouville Quantum Gravity, and the result is (depending on the correlation) either known or strongly suspected to yield a scale-invariant random metric with the topology of the $2$-sphere.
In $d=3$ the situation is, of course, much less clear, but our numerical study indicates that for the examined correlation matrices the distance profiles of $G_n^{\mathbf{C}}$ display accurate scaling with $n$.
Assuming this scaling persists to the full metric space and the Gromov--Hausdorff convergence of $n^{-1/d^{\mathbf{C}}_h}G_n^{\mathbf{C}}$ as $n\to\infty$ holds, this would establish a family of new universality classes of random geometries constructed from triples of correlated CRTs.
While the characteristics of the random geometries are yet to be studied in detail, two critical exponents of these prospective universality classes can be calculated or estimated from our data: the string susceptibility and the Hausdorff dimension. 

Our measurements pinpoint an interesting point on the boundary of parameter space where the off-diagonal elements of $\mathbf{C}$ approach $-1/2$, corresponding to a tiny equilateral spherical triangle ($\alpha=\beta=\gamma=\frac{\pi}{3}$), where the string susceptibility diverges ($\gamma_\mathrm{s}\rightarrow-\infty$) and the Hausdorff dimension appears to reach a minimum just above $4$.
This limit is analogous to the $\alpha\to 0$ limit of mating of trees in $d=2$, corresponding to the semi-classical limit $\gamma\rightarrow 0$ in 2-dimensional Liouville Quantum Gravity. 
Note that in both cases the covariance matrix $\mathbf{C}$ degenerates, and the Brownian motion effectively becomes $(d-1)$-dimensional, moving on the plane perpendicular to the diagonal.
However, since one is forcing the curve to perform a unit-time excursion in $\mathbb{R}_{\geq 0}^d$ the limit is rather singular, so it is not entirely clear how the classical $\gamma=0$ geometry is to be retrieved at $\alpha=0$ in $d=2$.
If one relaxes the positivity constraint in $d=2$, which naturally happens when consider infinite-volume limits, and considers Brownian motion that is nearly supported on the anti-diagonal in $\mathbb{R}^2$, the $\alpha\to 0$ limit leads to identifications of points at equal height in a single CRT, resembling the foliated structure of two-dimensional Causal Dynamical Triangulation \cite{ambjorn1998non} (see the final remarks of \cite{curien2019geometric}).
The analogous interpretation in the case $d=3$ amounts to the following. 
If we consider a two-dimensional Brownian motion on the $x+y+z=0$-plane in $\mathbb{R}^3$, then the first two components have covariance $\mathbf{C}_{12}=-1/2$.
Mating these two infinite correlated CRTs results in a random measure on $\mathbb{R}^2$ that is an infinite analogue of the unit-area $\gamma$-quantum sphere with $\gamma = \sqrt{4/3}$.
The third tree then leads to identification of certain pairs of points of $\mathbb{R}^2$ that have equal sum of heights within the two embedded trees.
It is a natural question to ask whether the discrete mating of trees bijection for bipolar-oriented triangulations, which lives in the $\gamma=\sqrt{4/3}$ universality class \cite{Kenyon_Bipolar_2019}, can incorporate such identifications to describe three-dimensional discrete geometries.

This picture extends to other points on the two-dimensional boundary of parameter space where $\alpha+\beta+\gamma = \pi$, but with the random metric on the plane replaced by that of Liouville Quantum Gravity with $\gamma = 2\sqrt{\alpha/\pi}$ and the identification performed by equal linear combination of the two heights (with coefficients $\sin \gamma$ and $\sin \beta$ respectively).
Here, as well as in the interior of the parameter space where $\det\mathbf{C} > 0$, one may ask the same question of whether discrete mating of trees bijections, like the ones in Section \ref{sec:stquad} and \ref{sec:perctri}, have a combinatorial interpretation at the level of discrete 3-manifolds. 
It would be preferable to take the opposite route, in which one starts with a combinatorial model of discrete 3-manifolds, like three-dimensional Dynamical Triangulations \cite{Ambjoern_Three_1992}, perhaps dressed with some matter statistical system and one would identify a bijective encoding into a triple of trees.
However, the combinatorics of discrete 3-manifolds is still poorly understood, making this a challenging route.
First steps towards encoding 3-sphere triangulations in trees has been taken by Lionni and one of the authors \cite{budd2022threespheres} by greatly restricting the type of triangulations considered.

In regard to Hausdorff dimension estimates using the numerical implementation of the Mated-CRT maps in the 2-dimensional case, our estimates for $\gamma =1,\sqrt{4/3},\sqrt{2},\sqrt{8/3}$ are statistically compatible with previous numerical results \cite{Barkley_2019} and rigorous bounds \cite{Ding:2018uez,gwynne2019bounds,Ang2019}. Moreover, this numerical toolbox proved to be reliable in sampling random geometries in the region $\gamma<1$ which has been inaccessible with other methods. We measured $d_H$ with good accuracy for $\gamma= 3/8, 1/2, 5/8, 3/4$. These results are compatible with a guessed formula of Ding and Gwynne, based on rigorous bounds \cite{Ding:2018uez}, and contradict Watabiki's formula, based on a heuristic heat kernel analysis in Liouville Quantum Gravity.

In the case $d=3$, a technical problems is finding sufficiently accurate numerical solutions to the harmonic equation \eqref{angular_dirichlet} in general cones, which could be further improved with the methods of \cite{dahne2020computation}. 
However, the main challenge in extending the results further out in the parameter space (and even to higher dimensions) is due to the large system sizes required, because of the following two reasons. 
As $1-\gamma_{\mathrm{s}} = \alpha_1$ becomes smaller, the distribution of the time extents of the Brownian motions $X^{1}(t)$ and $X^2(t)$ out of which we construct the excursion $\hat{X}(t)$ becomes increasingly heavy tailed (see Section \ref{sec:biasedbm}), making it harder to produce unbiased samples.
Secondly, the occurrence of higher Hausdorff dimensions means that a larger number of vertices is necessary to reach metric spaces of the same diameter, and this number is limited by the computing power available. 

Perhaps the most important question that we leave open in this work is whether the new family of scale-invariant random geometries, if it exists, describes anything resembling spacetime geometry, in particular whether it has manifold topology.
Deciding whether this is the case is considerably more difficult in $d=3$ compared to $d=2$.
One of the reasons is the lack of a natural interpretation of the Mated-CRT graphs as a discrete geometry of deterministic topology.
The other reason is that even if one has such a topology at the discrete level, there are many ways in which it can degenerate in the scaling limit.
In the two-dimensional case, there exist practical sufficient criteria that ensure the limit has $2$-sphere topology (see \cite{miermont2008sphericity} for a discussion and application to the Brownian sphere), while the situation in $d=3$ is less clear.

Short of answering these questions, having a catalogue of potential scale-invariant random geometries available is of value to research in Quantum Gravity.
It opens up the possibility of comparing characteristics of the UV fixed point in asymptotically safe gravity, established through other approaches, to the concrete list of models arising from mating of trees.
Drawing a bridge at the level of the dynamics is difficult, but a natural starting point is to compare critical exponents in various approaches. 
Hausdorff dimensions are often difficult to assess, since in approaches where the quantum geometry of spacetime is approximated with differentiable metrics, they tend to come out identical to the topological dimension. 
On the other hand, the string susceptibility, seen as the scaling behaviour of the partition function or as the distribution of sizes of minbus, should be easier to compare. Finally, the best studied critical exponent in Quantum Gravity appears to be the spectral dimension \cite{Reuter:2011ah,Calcagni:2013dna,Horava:2009if,Eichhorn2019SpectralDO}, which characterizes diffusion processes in the geometry. 
It is consistently found to decrease below the topological dimension in the UV. 
In the case of 2-dimensional mated-CRT maps, the spectral dimension is exactly equal to 2 for all $\gamma\in(0,2)$ \cite{Gwynne2017RandomWO}. 
A numerical estimation of the spectral dimension of the mated-CRT graphs in $d=3$ would be a logical follow up for the numerical methods developed in this work.


\end{document}